\documentclass{article}

\usepackage[margin=2cm]{geometry}
\usepackage{amsfonts}
\usepackage{graphicx} 
\usepackage{subfig}
\usepackage{cite}
\usepackage{bm}
\usepackage{caption}
\usepackage{comment}
\usepackage{amsmath}
\usepackage{amssymb}
\usepackage{color}
\usepackage{soul}
\usepackage{placeins}
\usepackage{tensor}
\usepackage[colorlinks=true,linktocpage=true,linkcolor=blue,citecolor=blue]{hyperref}
\usepackage[normalem]{ulem}
\usepackage{authblk}  
\usepackage{tikz}
\usepackage[utf8]{inputenc}
\usepackage{epsfig}
\usepackage{float} 
\usepackage[toc,page]{appendix}

\newcommand{\rg}[1]{\textcolor{blue}{[RG: #1]}}
\newcommand{\RG}[1]{{\bf{\textcolor{blue}{RG: #1}}}}

\title{Boson Star Factory: Past, Present, and Future}
\author[1,2]{Romain Gervalle\footnote{romaingervalle@ua.pt}}
\author[1]{Jorge F. M. Delgado\footnote{jorgedelgado@ua.pt}}
\author[1,3]{Etevaldo dos Santos Costa Filho\footnote{etevaldo.s.costa@ua.pt}}
\author[1,4]{Carlos A. R. Herdeiro\footnote{herdeiro@ua.pt}}
\author[5]{Víctor Jaramillo\footnote{victor.jaramillo@academicos.udg.mx}}
\author[6, 7]{Raimon Luna\footnote{raimonluna@gmail.com}}
\author[1]{Eugen Radu\footnote{eugen.radu@ua.pt}}

\affil[1]{\normalsize Departamento de Matemática da Universidade de Aveiro and 

Center for Research and Development in Mathematics and Applications -- CIDMA

Campus de Santiago, 3810-183 Aveiro, Portugal}

\affil[2]{Département de Physique de l'Université de Tours, Parc de Grandmont, 37200 Tours, France}

\affil[3]{Programa de Pós-Graduação em Física, Universidade Federal do Espírito Santo, 
Vitória, ES,  29075-910, Brazil}

\affil[4]{Programa de P\'os-Gradua\c{c}\~{a}o em F\'{\i}sica, Universidade Federal do Par\'a, 66075-110, Bel\'em, Par\'a, Brazil}

\affil[5]{\normalsize Departamento de Física, Centro Universitario de Ciencias Exactas e Ingenierías,
Universidad de Guadalajara, 44430 Guadalajara, Jalisco, México}

\affil[6]{Departament de F\'\i sica Qu\`antica i Astrof\'\i sica,  Universitat de Barcelona,
Mart\'\i\  i Franqu\`es 1, \mbox{ES-08028}, Barcelona, Spain.}

\affil[7]{Institut de Ci\`encies del Cosmos (ICCUB),  Universitat de Barcelona,
Mart\'\i\  i Franqu\`es 1, \mbox{ES-08028}, Barcelona, Spain.}

\begin{document}

\date{\today}

\maketitle

\begin{abstract}
    Boson stars are self-gravitating solitons of the Einstein-Klein-Gordon equations for a massive complex scalar field, and arguably the simplest horizonless compact objects arising in General Relativity. Beyond spherical symmetry, their construction requires solving a system of nonlinear elliptic partial differential equations, a difficult task \textit{per se} and that each group has traditionally addressed with its own numerical implementation. A systematic comparison of different methods applied to the same physical model has so far been lacking. In this work, we carry out such a comparison for the two simplest non-spherical configurations: rotating boson stars and static dipolar boson stars. We employ three independent codes, based respectively on finite differences (\texttt{FIDISOL/CADSOL}), finite elements (\texttt{FreeFem}), and spectral methods (\texttt{Kadath}), and assess their accuracy through different diagnostics. The three solvers agree on the global observables to typically 
    eight significant 
    digits. We further compare the runtime of each code on identical hardware and provide precise benchmark values of the relevant physical quantities, intended to serve as reference data for future numerical studies of boson stars. As a further proof of concept, we also present the use of physics-informed neural networks to construct a rotating $Q$-ball solution.
\end{abstract}

\tableofcontents
\section{Introduction}
 
The simplest extension of electro-vacuum General Relativity (GR) is arguably scalar-vacuum.
Allowing the scalar field to be massive and complex, but still free and minimally coupled to gravity, leads to a model with a rich spectrum of solutions, some of them with no counterparts in the electro-vacuum case.

Of interest in the context of this work are 
the Boson Stars (BSs).
These self-gravitating  solitons composed of a massive complex field
were first introduced by Kaup \cite{Kaup:1968zz} as well as Ruffini and Bonazzola \cite{Ruffini:1969qy} almost 60 years ago (see also Feinblum and McKinley~\cite{Feinblum:1968nwc}).
The  BSs are globally regular, asymptotically flat, equilibrium solutions
of the Einstein--Klein-Gordon (EKG) equations, 
providing an explicit realization of Wheeler's geons 
\cite{Wheeler:1955zz}.
They are generally regarded as {\it `macroscopic quantum states'},    
which are prevented from
gravitational collapse by Heisenberg's uncertainty principle.

After a few decades of rather limited interest,
in recent years, BSs have received increasing 
attention in a variety of contexts. 
They have been proposed $e.g.$ as black-hole mimicker (as in the early study 
\cite{Guzman:2009zz})
or as candidate objects for dark matter haloes 
\cite{Sin:1992bg}
(see \cite{Liebling:2012fv} for a review of these aspects).

A major obstacle in the study of BSs is the absence of exact solutions, with all known solution-generating techniques failing in the presence of a mass term for the scalar field. For this reason, the earlier results \cite{Kaup:1968zz,Ruffini:1969qy} have primarily focused on spherically symmetric configurations whose construction is relatively straightforward, requiring to solve numerically a system of nonlinear ordinary differential equations (ODEs). While spinning generalizations were anticipated early on, rotating BSs present an additional difficulty: unlike neutron stars or Kerr black holes, the solutions do not admit a slowly rotating limit~\cite{Kobayashi:1994qi} and must therefore be constructed directly in the fully rotating regime. This requires solving a complicated system of nonlinear partial differential equations (PDEs), making their numerical construction considerably more involved.

The first construction of rotating BSs (RBSs) was reported in the literature by Schunck and Mielke \cite{Schunck:1996he}, who showed that BSs can rotate only differentially, not uniformly. 
This result has been subsequently reproduced and extended by many authors, using different numerical methods.
We briefly review  below the most relevant work, with emphasis on the numerical approach.
First, let use remark that  the results in \cite{Schunck:1996he} were restricted to individual solutions in the weakly relativistic regime, with the value $m=500$ of the rotational quantum number in the scalar field harmonic azimuthal dependence, $cf.$
eq. (\ref{scalar_ansatz}). Also, the  Schunck-Mielke solutions were found by directly solving the EKG equations as a boundary value problem using finite differences. 

Shortly thereafter, Yoshida and Eriguchi~\cite{Yoshida:1997qf} constructed for the first time highly relativistic RBSs using an integral representation of the EKG equations and the self-consistent-field method (see Ref.~\cite{Yoshida:1997nd} and references therein for more details about this approach). They computed full sequences of solutions with $m=1,2$, and provided numerical values for the maximal mass and Noether charge in the $m=1$ case. 

In 2005, Kleihaus \textit{et al.} \cite{Kleihaus:2005me} constructed RBSs with a sextic self-interaction potential for azimuthal number $m=1$, reporting solutions even in the numerically demanding higher branches of the mass-frequency diagram. Their numerical approach is based on finite differences. Another study of RBSs has then been reported by Grandclément, Somé and Gourgoulhon \cite{Grandclement:2014msa} who considered both free and self-interacting scalar fields (with quartic and sextic potentials). The EKG equations are solved via spectral methods, and their results confirmed those of Yoshida and Eriguchi for $m=1$, provided the maximal mass for free-field BSs with $m=2$, and presented the first construction of rotating solutions with $m=3$ and $m=4$. These results were further confirmed and extended by Ontañón and Alcubierre \cite{Ontanon:2021hbg} (using a
finite difference solver),
who provided results for the maximal mass and minimal frequency for RBSs up to $m=6$.

RBSs are axially symmetric, having of a toroidal energy density distribution; however, rotation is not the only mechanism to evade spherical symmetry. The first \textit{static} non-spherical BSs were reported by Schupp and van der Bij~\cite{Schupp:1995dy}, whose configurations describe two symmetric lumps of scalar energy density in static equilibrium, with gravitational attraction counterbalanced by scalar repulsion. The solutions were obtained using a finite element solver -- one of the earliest applications of this approach in the boson star literature, albeit within the Newtonian approximation. These dipolar BSs (DBSs) were subsequently extended to the full GR framework by Yoshida and Eriguchi \cite{Yoshida:1997nd},
using the same approach as in Ref.~\cite{Yoshida:1997qf}, and providing an estimation of the maximal mass.
 These results were confirmed and extended by Cunha \textit{et al.} \cite{Cunha:2022tvk} using a finite differences solver.
 In a related development, Herdeiro \textit{et al.}~\cite{Herdeiro:2020kvf} have reported more complex configurations consisting of several scalar lumps in equilibrium, dubbed multipolar BSs. The simplest among these represent chains of BSs, being static and axially symmetric with several distinct components located along the symmetry axis. Static chains were studied specifically in Ref.~\cite{Herdeiro2021}, and their rotating generalizations in Ref.~\cite{Gervalle2022}, both in the case of a sextic self-interaction potential.

To summarize, the study of axially symmetric BSs requires to use robust numerical methods to solve the underlying elliptic field equations. Historically, finite differences were the first method to be employed for non-spherical BSs, an approach which remains widely used in the strong-gravity community.
 A representative solver here is  \texttt{FIDISOL/CADSOL}, which has been employed to found both RBSs  \cite{Kleihaus:2005me}  and DBSs  \cite{Cunha:2022tvk}, as well as a variety of other solutions in non-linear field theories.
The use of spectral and finite element methods is more recent, and has yet to achieve widespread adoption in the field. Representative solvers in these categories are the \texttt{Kadath} spectral library~\cite{Grandclement:2009ju} and the \texttt{FreeFem} finite element library~\cite{MR3043640}, which have only recently been applied to the construction of axisymmetric BSs: \texttt{Kadath} was first employed for RBSs in 2014~\cite{Grandclement:2014msa}, and later for DBSs in 2025~\cite{Jaramillo:2024cus}, while the application of \texttt{FreeFem} to strong-gravity problems was essentially absent from the literature prior to Ref~\cite{Gervalle2022}. Nevertheless, spectral methods more broadly, including \texttt{Kadath}'s precursor \texttt{LORENE}, have a much longer history in strong-gravity.
  
  At the same time, a systematic comparison of results obtained from different solvers applied on the same physical model
  is still lacking 
  in the literature, as each group typically relies on its own numerical implementation. 
  The main purpose of the present work is precisely to carry out such a comparison in a systematic and quantitative manner, assessing whether the three aforementioned approaches yield consistent results. BSs provide a natural testbed for this comparison, as they constitute arguably the simplest class of horizonless compact objects arising in a gravitating field theory. We further restrict to rotating and dipolar BSs, which are the simplest instance of non-spherical configurations in the EKG model.

\medskip

We shall consider three numerical codes based respectively on $i)$ finite differences (\texttt{FIDISOL/CADSOL}), $ii)$ finite elements (\texttt{FreeFem}), and $iii)$ spectral methods (\texttt{Kadath}).
As benchmark configurations, we consider three $m=1$ RBSs and three DBSs. Since the internal convergence criteria differ substantially among the three numerical frameworks, they cannot serve as a common measure of accuracy. We therefore base the comparison primarily on physical, code-independent diagnostics, 
the most important being the relative differences between the Komar and ADM definitions of mass and angular momentum: the most basic physical observables accessible to gravitational wave detectors. As an additional error estimator, we also introduce a virial identity: an integral whose value is zero on-shell by construction, yet whose integrand is nonvanishing. For the six reference configurations, we also provide high-precision values of key physical quantities by pushing each solver to its limits of achievable accuracy and comparing their agreement. Additionally, the runtime of each code is compared on the same hardware. All these results may serve as benchmark data for the development of future numerical frameworks, particularly in the context of artificial intelligence-based methods such as physics-informed neural networks (PINNs).

\medskip

The paper is organized as follows. In Sec.~\ref{sec:general_aspects}, we introduce the model along with the axially symmetric field ansatz, reduced equations of motion, boundary conditions and relevant physical quantities. The three numerical methods and their corresponding implementations are described in Sec.~\ref{sec:methods_software}, while general properties of the BS families are discussed in Sec.~\ref{sec:gen_properties}. The main results are collected in Sec.~\ref{sec:results}, including the detailed comparison of the different solvers on six reference solutions, convergence tests, and a runtime benchmark. This section also contains some technical details specific to each solver. Future prospects, in particular the use of PINNs, are discussed in Sec.~\ref{sec:future}. We finally summarize our results and provide concluding remarks in Sec.~\ref{sec:conclusion}.

\section{Boson Stars: general aspects}
\label{sec:general_aspects}

\subsection{The model}

\medskip

We shall be working with the EKG field theory, describing a massive complex scalar field 
$\Psi$ minimally coupled to Einstein gravity. The model has the following action,
\begin{equation}
\label{action}
S=\int  d^4x \sqrt{-g}\left[ \frac{R}{16\pi G}
   -\frac{1}{2} g^{ab}\left( \Psi_{, \, a}^* \Psi_{, \, b} + \Psi _
{, \, b}^* \Psi _{, \, a} \right) - \mu^2 \Psi^*\Psi
 \right],
\end{equation}
where $G$ is Newton's constant and $\mu$, the scalar field mass.
The corresponding EKG  field equations, obtained from the variation of the action with respect to the metric and scalar field 
read, respectively,
\begin{eqnarray}
\label{E-eq}
&&
E_{ab}\equiv R_{ab}-\frac{1}{2}g_{ab}R-8 \pi G~T_{ab}=0 \, ,
 \\
&&
\label{KG-eq}
\Box \Psi=\mu^2\Psi \ ,
\end{eqnarray}  
where,
\begin{equation}
    T_{ab}\equiv  
 \Psi_{ , a}^*\Psi_{,b}
+\Psi_{,b}^*\Psi_{,a} 
-g_{ab}  \left[ \frac{1}{2} g^{cd} 
 ( \Psi_{,c}^*\Psi_{,d}+
\Psi_{,d}^*\Psi_{,c} )+\mu^2 \Psi^*\Psi\right],  
\end{equation}
is the energy-momentum tensor of the scalar field.

The action (\ref{action}) is invariant under the global $U(1)$ transformation $\Psi\rightarrow e^{i\alpha}\Psi$, where $\alpha$ is constant. Thus, the scalar 4-current, $j^a=-i (\Psi^* \partial^a \Psi-\Psi \partial^a \Psi^*)$, is conserved:  $j^a_{\ ;a}=0$. It follows that integrating the timelike component of this 4-current over a spacelike slice $\Sigma$ yields a conserved quantity -- the \textit{Noether charge}:
\begin{eqnarray}
\label{Q}
Q=\int_{\Sigma}d^3 x\sqrt{-g}\,j^t \ .
\end{eqnarray}
At a microscopic level, this Noether charge counts the number of scalar particles. 

\medskip

\subsection{The Ansatz  }
\label{ansatz}

The axially symmetric BSs
can obtained by using the following metric ansatz\footnote{A reparametrization of (\ref{metric_ansatz}) 
is employed sometimes in the literature, with
$e^{2F_0}=f$, 
$e^{2F_1}=lg /f$,
$e^{2F_2}=l  /f$
and 
$W=\Omega/r$
(with $f,g,l,\Omega$
functions of $r,\theta$).
Although
this alternative metric formulation may possess some advantages 
in the numerical treatment of the problem, it leads to a more complicated form of the equations of motion than those displayed in this work.

Also, 
the \texttt{FIDISOL/CADSOL} solver uses a form of  the line element (\ref{metric_ansatz}) with 
$W=\Omega/r$, which allows for a direct computation of the total angular momentum.
The function $\Omega(r,\theta)$ is subject to the same boundary conditions as $W(r,\theta)$.
} 
with four unknown functions of the radial coordinate $r$ and polar angle $\theta$,
\begin{eqnarray}
\label{metric_ansatz}
ds^2=-e^{2F_0(r,\theta)}   dt^2+e^{2F_1(r,\theta)}\left({dr^2}+r^2 d\theta^2\right)+e^{2F_2(r,\theta)}r^2 \sin^2\theta \left(d\varphi-W(r,\theta) dt\right)^2 \ . 
\end{eqnarray}
This ansatz is written in the \textit{quasi-isotropic} coordinates $(t,r,\theta,\varphi)$ and the absence of the cross terms $g_{tr}$, $g_{t\theta}$, $g_{\varphi r}$ and $g_{\varphi\theta}$ amounts to assuming a \textit{circular} spacetime. 

The ansatz for the scalar field
reads,
\begin{eqnarray}
\Psi=\phi(r,\theta)e^{i(m\varphi- \omega t)}~,
\label{scalar_ansatz}
\end{eqnarray} 
 where $\phi$, a real function, is the scalar field amplitude, $\omega>0$ is the scalar field frequency and $m$, an integer, is the azimuthal harmonic index. The form of the scalar field is motivated by well-known no-go arguments: the scaling argument due to Derrick \cite{Derrick1964} for flat-space solitons, and the integral identity due to Bekenstein \cite{Bekenstein:1972ny} originally formulated for black holes. Although derived in different contexts, both arguments can be adapted to the globally regular, horizonless, gravitating configurations of interest here. In brief, a scalar field inheriting the symmetries of a stationary and axially symmetric spacetime ($i.e.$ being independent of both $t$ and $\varphi$) cannot source any non-trivial configuration: integrating the field equation over a spacelike slice, the only solution consistent with regularity and asymptotic flatness is the trivial vacuum. Choosing the harmonic\footnote{Note that the model~\eqref{action} with one complex scalar field is equivalent to a model containing two real scalar fields, for which an ansatz equivalent to \eqref{scalar_ansatz} can be derived by choosing appropriate periodic time and azimuthal dependencies for each real component.} time and azimuthal dependence as in \eqref{scalar_ansatz} is a natural way to evade this conclusion while preserving the targeted spacetime symmetries as the energy-momentum tensor remains stationary and axisymmetric even though the scalar field itself is not. Note that the circularity property of the line element \eqref{metric_ansatz} is consistent with the scalar field ansatz \eqref{scalar_ansatz}.
 
The value of $m$ together with the behaviour of the scalar field 
under a reflection in the equatorial plane $\theta=\pi/2$
defines the two classes of 
solutions we shall consider. 
These are
\begin{itemize}
    \item 
   \underline{{\it  The rotating Boson Stars  (RBSs):} }$m=\pm 1, \pm 2 \dots$ in (\ref{scalar_ansatz});
  there  we shall restrict our study to the  case of a scalar field amplitude  $\phi$ 
    which does not change sign $w.r.t.$ a reflection symmetry along the equatorial plane,
     $\phi(r, \theta)= \phi(\pi -\theta)$.
        \item 
     \underline{{\it  The static dipolar Boson Stars (DBSs):}} $m=0$ in (\ref{scalar_ansatz})
     and $W(r,\theta)=0$ in (\ref{metric_ansatz}).
     In this case the $\phi$-function 
     is antisymmetric \textit{w.r.t.} the equatorial plane,
 $\phi(r, \theta) = -\phi(r,\pi -\theta)$.
\end{itemize}

\medskip
Finally, let us remark that
 the metric ansatz (\ref{metric_ansatz})
possesses two Killing vector fields,
\begin{equation}
  \xi=\partial_t, \qquad  {\rm and} \qquad \eta=\partial_\varphi.  
  \label{eq:killings}
\end{equation}
which, however, are $not$ symmetries of the full solution, since they do not preserve the expression of the scalar field. The only symmetry of the full solution is generated by the helicoidal vector field (for $m\neq 0$),
  \begin{equation} 
  \chi =\xi+\frac{\omega}{m} \eta,
  \label{helicoidal}
  \end{equation}
since $\chi^\mu\, \partial_\mu\Psi=0$. For DBSs such symmetry does not apply.

\subsection{The equations of motion }
\label{eom}
 
As we previously mentioned, we are interested in stationary solutions of the EKG field equations. For the numerical construction of such solutions, it is important to have a well-posed boundary value problem, which in practice means casting the field equations into an elliptic system. Let $\Omega\subset\mathbb{R}^2$ be the effective computational domain with coordinates,
\[
        X^i=(X,Y)=(r,\theta),\qquad i=1,2,
\]
and let,
\[
        \mathcal F=(\mathcal F^1,\ldots,\mathcal F^{N_K})
        :\Omega\to\mathbb{R}^{N_K},
\]
denote the vector of unknown fields. A generic second-order PDE system can be written as,
\begin{equation}
\label{eq:cadsol-operator}
        \mathcal E_A
        \left(X^i;\mathcal F,\mathcal F_{,i},\mathcal F_{,ij}\right)=0,
        \qquad
        A=1,\ldots,N_{K},\qquad i,j=1,2 ,
\end{equation}
where each equation of motion, \(\mathcal E_A\), may depend nonlinearly on the fields and on their derivatives up to second order. 
The principal coefficients of the system are \cite{Giaquinta1984MultipleII,giaquinta2012introduction,Koshelev1995},
\begin{equation}
\label{eq:principal-coefficients}
        \mathcal A^{ij}{}_{AB}
        =
        \frac{\partial \mathcal E_A}{\partial \mathcal{F}^B{}_{,mn}},
        \qquad
        A,B=1,\ldots,N_{K},\qquad i,j=1,2.
\end{equation}
Note that for sufficiently regular fields, the mixed second derivatives commute, $\mathcal{F}_{,XY}=\mathcal{F}_{,YX}$, and the above coefficients are symmetric in $(i,j)$.
For quasilinear equations, the system is said to be elliptic if \cite{Giaquinta1984MultipleII,giaquinta2012introduction,Koshelev1995},

\begin{equation}
\label{eq:ellipticity}
       \forall\,\zeta=(\zeta_1,\zeta_2)\in\mathbb R^2,\quad
        \forall\,q=(q^1,\ldots,q^{N_K})\in\mathbb R^{N_K},\qquad  \mathcal A^{ij}{}_{AB}\,\zeta_i\zeta_j\,q^Aq^B
        >0.
\end{equation}

Even though many stationary systems in GR and modified gravity reduce to \emph{coupled, second-order, nonlinear elliptic} PDEs after symmetry reduction \cite{Besse2007}, 
the elliptic nature of Einstein's equations is not necessarily manifest, 
and a direct numerical construction of stationary solutions is therefore difficult. To see this explicitly, rewrite Einstein's equations in \eqref{E-eq} as,
 \begin{equation}
 		R_{ab}=8\pi G \left(T_{ab}-\dfrac{1}{2}g_{ab}T\right)\,.\\
        \label{eq:ein_alt}
 \end{equation}
 %
 Then, the principal part of the differential operator 
 is contained in the Ricci tensor,
\begin{equation}
	R_{ab}
	= R^{c}{}_{acb}
	= \partial_{c}\Gamma^{c}{}_{ab}
	- \partial_{a}\Gamma^{c}{}_{cb}
	+ \Gamma^{c}{}_{ab}\Gamma^{d}{}_{cd}
	- \Gamma^{c}{}_{db}\Gamma^{d}{}_{ca},
\end{equation}
where $\Gamma^{a}{}_{bc}$ are the Christoffel symbols. To determine the character of the resulting system of differential equations, we need to examine the
principal coefficients of the differential operator. Hence, making the second derivatives of the metric explicit, one can rewrite $R_{ab}$,
\begin{equation}
	\begin{aligned}
		R_{ab}
		= -\,\frac{1}{2} g^{cd}\left(
		\partial_c \partial_d g_{ab}
		+ \partial_a \partial_b g_{cd}
		- \partial_a \partial_c g_{bd}
		- \partial_b \partial_c g_{ad}\right)
		\\[2pt]
		\quad
		+\, g^{cd} g_{ef}\left(
		\Gamma^{e}{}_{ca} \Gamma^{f}{}_{db}
		- \Gamma^{e}{}_{ab} \Gamma^{f}{}_{cd}\right)\, .
	\end{aligned}
    \label{eq:ricci}
\end{equation}

The first term in this expression is of Laplace type and has the desired elliptic principal part \cite{Choquet-Bruhat:2009xil}. However, the other terms in the first line also contain second derivatives of the metric and therefore modify the principal symbol (see Section IV of the review \cite{Dias:2015nua} for the explicit form of the operator at the linear level). A common strategy is to modify or reorganize the equations so that one obtains an equivalent elliptic set of equations together with additional constraint relations. The latter being the strategy we shall consider in this paper.

Specifically, choosing the coordinate system as in \eqref{metric_ansatz}, the combinations of the Einstein equations \eqref{E-eq} yielding an elliptic set of PDEs are\footnote{For the \texttt{FreeFem} solver, the combination in Eq.~\eqref{eq:elliptic-F1} is replaced by $\tensor{E}{^r_r}-\tensor{E}{^\theta_\theta}+\tensor{E}{^\varphi_\varphi}+\tensor{E}{^t_t}=0$. This alternative combination is equivalent to the original one, as it differs from it only by $2\tensor{E}{^r_r}$, but has a cleaner spherically symmetric limit, in which it reduces directly to Eq.~\eqref{eq:elliptic-F2} when $F_1=F_2$. This choice was found to give better numerical behavior near the symmetry axis in \texttt{FreeFem}.},

\begin{align}
        \mathcal E_{F_1}
        &\equiv
        \frac{r^2 e^{2F_1}}{2}
        \left(
        -E^r{}_r-E^\theta{}_\theta+E^\varphi{}_\varphi+E^t{}_t
        \right)=0,                                      \label{eq:elliptic-F1}
        \\
        \mathcal E_{F_2}
        &\equiv
        \frac{r^2 e^{2F_1}}{2}
        \left(
        E^r{}_r+E^\theta{}_\theta-E^\varphi{}_\varphi+E^t{}_t
        +2W E^t{}_\varphi
        \right)=0,                                      \label{eq:elliptic-F2}
        \\
        \mathcal E_{F_0}
        &\equiv
        \frac{r^2 e^{2F_1}}{2}
        \left(
        E^r{}_r+E^\theta{}_\theta+E^\varphi{}_\varphi-E^t{}_t
        -2W E^t{}_\varphi
        \right)=0,                                      \label{eq:elliptic-F0}
        \\
        \mathcal E_W
        &\equiv
        -\frac{2e^{2(F_0+F_1-F_2)}}{\sin^2\theta}
        E^t{}_\varphi=0 .                                  \label{eq:elliptic-W}
\end{align}
 The remaining independent
Einstein equations can be chosen as,
\begin{equation}
  E^{r}{}_{\theta} = 0,
  \qquad
  E^{r}{}_{r} - E^{\theta}{}_{\theta} = 0\label{eq:cons}\,,
\end{equation}
which we regard as \emph{constraint equations}. These constraints are not solved directly in our numerical schemes, but they must be satisfied by any consistent solution.
In fact
one can show that, given our choice of elliptic equations, the constraints \(E^{r}{}_{\theta} = 0\) and \(E^{r}{}_{r} - E^{\theta}{}_{\theta} = 0\)
are automatically obeyed once we impose the boundary conditions \cite{Herdeiro:2015gia,Wiseman:2002zc}. In practice, we thus solve
only the elliptic equations \eqref{eq:elliptic-F1}-\eqref{eq:elliptic-W} (together with
the 
scalar field equation) and we monitor the residuals of the two constraints \textit{a posteriori}.

Within this framework, the EKG field equations are thus reduced to a set of five
coupled, nonlinear, elliptic partial differential equations (PDEs) for the
functions ${\cal F}=(\phi,F_1,F_2,F_0,W)$ (four for the static dipolar stars,
since $W\equiv 0$ in that case). The first of these is the Klein-Gordon equation in
\eqref{KG-eq}, whose explicit form reads
\begin{eqnarray}
\label{eq-phi}
&&
r^2\phi_{,rr}+\phi_{,\theta\theta}
+r^2\phi_{,r}(F_{0,r}+F_{2,r})
+\phi_{,\theta}(F_{0,\theta}+F_{2,\theta})
+2r\phi_{,r}+\cot\theta\,\phi_{,\theta}
\\
\nonumber
&&
{~~~~~~~~}
-
\left(
\frac{e^{-2F_2}m^2}{\sin^2\theta}
-r^2e^{-2F_0}(\omega-mW)^2
+r^2\mu^2
\right)
e^{2F_1}\phi=0\ .
\end{eqnarray}
For the metric functions, the four combinations \eqref{eq:elliptic-F1}--\eqref{eq:elliptic-W} yield the following elliptic equations,
\begin{eqnarray}
\label{eq-F1}
&&
r^2F_{1,rr}+F_{1,\theta\theta}
-\left(r^2F_{0,r}F_{2,r}+F_{0,\theta}F_{2,\theta}\right)
-\frac{e^{-2F_0+2F_2}r^2\sin^2\theta}{4}
\left(r^2W_{,r}^2+W_{,\theta}^2\right)
-rF_{0,r}
\\
\nonumber
&&
{~}+rF_{1,r}
-\cot\theta\,F_{0,\theta}
+8\pi G
\left(
r^2\phi_{,r}^2+\phi_{,\theta}^2
+e^{2F_1}
\left[
r^2e^{-2F_0}(\omega-mW)^2
-\frac{e^{-2F_2}m^2}{\sin^2\theta}
\right]\phi^2
\right)
=0\ ,
\end{eqnarray}
\begin{eqnarray}
\label{eq-F2}
&&
r^2F_{2,rr}+F_{2,\theta\theta}
+r^2F_{2,r}^2+F_{2,\theta}^2
+r^2F_{0,r}F_{2,r}+F_{0,\theta}F_{2,\theta}
+\frac{e^{-2F_0+2F_2}r^2\sin^2\theta}{2}
\left(r^2W_{,r}^2+W_{,\theta}^2\right)
\\
\nonumber
&&
{~}+rF_{0,r}+\cot\theta\,F_{0,\theta}
+3rF_{2,r}+2\cot\theta\,F_{2,\theta}
+8\pi G\,e^{2F_1}
\left(
r^2\mu^2+\frac{2e^{-2F_2}m^2}{\sin^2\theta}
\right)\phi^2
=0\ ,
\end{eqnarray}
\begin{eqnarray}
\label{eq-F0}
&&
r^2F_{0,rr}+F_{0,\theta\theta}
+r^2F_{0,r}^2+F_{0,\theta}^2
+r^2F_{0,r}F_{2,r}+F_{0,\theta}F_{2,\theta}
-\frac{e^{-2F_0+2F_2}r^2\sin^2\theta}{2}
\left(r^2W_{,r}^2+W_{,\theta}^2\right)
\\
\nonumber
&&
{~}+2rF_{0,r}
+\cot\theta\,F_{0,\theta}
-8\pi G\,r^2e^{2F_1}
\left[
2e^{-2F_0}(\omega-mW)^2-\mu^2
\right]\phi^2
=0\ ,
\end{eqnarray}
\begin{eqnarray}
\label{eq-W}
&&
r^2W_{,rr}+W_{,\theta\theta}
+r^2(3F_{2,r}-F_{0,r})W_{,r}
+(3F_{2,\theta}-F_{0,\theta})W_{,\theta}
\\
\nonumber
&&
{~}
+4r\left(
W_{,r}+\frac{3\cot\theta\,W_{,\theta}}{4r}
\right)
+32\pi G\,
\frac{e^{2F_1-2F_2}m(\omega-mW)}{\sin^2\theta}
\phi^2
=0\ .
\end{eqnarray}
One observes that the above equations all share the desired Laplace-like
principal part: $\,r^{2}\,\partial_{r}^{2} \mathcal{F}
	+ \,\partial_{\theta}^{2} \mathcal{F}$.
The corresponding system for the static DBSs follows by setting
$W=m=0$ in  \eqref{eq-phi}-\eqref{eq-W}. The three numerical schemes we shall use to solve these systems of PDEs are described in Sec.~\ref{sec:methods_software} below (additional solver-specific details are also given in Sec.~\ref{sec:specs}).

For completeness, we also give the explicit form of the two constraint equations \eqref{eq:cons} 
\begin{eqnarray}
\nonumber
&&
F_{0,rr}-\frac{1}{r^2 }  F_{0,\theta\theta} 
+F_{2,rr}-\frac{1}{r^2 }  F_{2,\theta\theta} 
+F_{0,r}^2-\frac{1}{r^2 }  F_{0,\theta}^2
-2\left(
F_{0,r}F_{1,r}-\frac{1}{r^2 }F_{0, \theta}  F_{1,\theta} 
\right) 
\\
\nonumber
&&
{~}-2\left(
F_{1,r}F_{2,r}-\frac{1}{r^2 }F_{1, \theta}  F_{2,\theta} 
\right)
-\frac{e^{-2F_0+2F_2}r^2\sin^2\theta}{2 }
\left(
W_{,r}^2-\frac{1}{r^2  }W_{, \theta}^2
\right)
+
F_{2,r}^2-\frac{1}{r^2  }F_{2, \theta}^2
\\
\nonumber
&&
{~}-\frac{1}{r}\left(F_{0,r}+2F_{1,r}-F_{2,r}\right)
+\frac{2\cot\theta}{r^2 }
(F_{1,\theta}-F_{2,\theta})
+16\pi G 
\left(\phi_{,r}^2-\frac{1}{r^2 }\phi_{, \theta}^2\right)
=0\ ,
\end{eqnarray}
and
\begin{eqnarray}
\nonumber
&&
 F_{0,r\theta}+
 F_{2,r\theta}+
 F_{0,r} F_{0,\theta}+
 F_{2,r} F_{2,\theta}-
 ( F_{0,r} F_{1,\theta}+
 F_{1,r} F_{0,\theta})
 - ( F_{1,r} F_{2,\theta}+
 F_{2,r} F_{1,\theta})
 \\
 \nonumber
 &&
{~} -\frac{F_{0,\theta}}{r}
  -\frac{F_{1,\theta}}{r}
 -\cot\theta(F_{1,r}-F_{2,r})
 -\frac{e^{-2F_0+2F_2}r^2\sin^2\theta}{2 } W_{,r} W_{,\theta}
 +16\pi G  \phi_{,r} \phi_{,\theta}=0\ .
\end{eqnarray}
 %

\subsection{Boundary conditions  }
\label{boundary}


The unknown functions $\cal F$ depend on $(r,\theta)$ with $r\in[0,\infty)$ and $\theta\in[0,\pi]$. Requiring the energy density to be symmetric $w.r.t.$ the equatorial plane allows us to restrict the computational domain to $(r,\theta)\in[0,\infty)\times[0,\pi/2]$. We now specify the boundary conditions at infinity $r\to\infty$, at the origin $r=0$, on the symmetry axis $\theta=0$, and on the equatorial plane $\theta=\pi/2$.

We restrict our attention to asymptotically flat configurations, for which the geometry approaches the Minkowski 
spacetime as $r\to\infty$. For our purposes, the line element \eqref{metric_ansatz} is
said to be asymptotically flat if, at large $r$, it approaches the Minkowski
metric with the following decay,
\begin{equation}
	\label{AF-expansion}
	\begin{aligned}
		ds^2 ={}&
		- \left( 1 - \frac{2GM}{r} +\mathcal{O}(r^{-2}) \right) dt^{2}
		- \left( \frac{4GJ}{r}\sin^{2}\theta + \mathcal{O}(r^{-2}) \right) dt\,d\varphi
		\\[2mm]
		&\quad
		+ \left( 1 + \mathcal{O}(r^{-1}) \right)
		\Bigl[ dr^{2}
		+ r^{2}\bigl( d\theta^{2}
		+ \sin^{2}\theta\, d\varphi^{2} \bigr)
		\Bigr] .
	\end{aligned}
\end{equation}
Here $M$ and $J$ are, respectively, the ADM mass and angular momentum of
spacetime. Therefore, the metric functions in \eqref{metric_ansatz} satisfy the boundary conditions at infinity,
\begin{equation}\label{bc_inf}
	F_i|_{r\to\infty}= W|_{r\to\infty}=0.
\end{equation}
Finite energy requires the scalar field to be spatially localized with an exponential decay at large distance and the bound state condition $\omega<\mu$. The corresponding boundary condition at infinity is, 
 \begin{equation}\label{bc_inf1}
\phi\big|_{r\to\infty}=0.
 \end{equation}
 

\medskip

Requiring regularity at the origin and inspecting the small-$r$ expansion of the equations \eqref{eq-phi}--\eqref{eq-W} lead to
the following boundary conditions at $r=0$ :
\begin{equation}
\label{r=0}
\partial_r F_i\big|_{r=0}=\partial_r W\big|_{r=0} =0,\quad \phi\big|_{r=0}=0.
\end{equation}
The condition $\phi(0)=0$ is imposed for both RBS and DBS families, but it originates from genuinely different mechanisms. For RBSs ($|m|\geq1$), the local expansion analysis of \eqref{eq-phi} fixes the leading behaviour $\phi\sim(r\sin\theta)^{|m|}$, so that the scalar amplitude vanishes on the axis
and at the centre, before growing away from the axis. For the static DBSs ($m=0$) there is no centrifugal barrier, and $\phi(0)=0$ follows instead from the equatorial antisymmetry $\phi(r,\theta)=-\phi(r,\pi-\theta)$: the origin lies in the equatorial plane $\theta=\pi/2$, where the amplitude vanishes by the odd parity requirement. In this case, the local expansion analysis yields $\phi\sim r\cos\theta$ so that the scalar amplitude vanishes at the centre but grows along the symmetry axis. By contrast, a spherically symmetric BS would have $\phi(0)\neq0$.

\medskip

Let $X \equiv \eta\cdot\eta$ denote the squared norm of the axial Killing
vector $\eta$ in \eqref{eq:killings}. Once the azimuthal coordinate $\varphi$ is fixed to have
period $2\pi$, regularity of the geometry on the symmetry axis requires the absence of conical singularities at
the poles. This requirement is equivalent to imposing, on the axis
\cite{stephani_kramer_maccallum_hoenselaers_herlt_2003},
\begin{equation}\label{conical_generic}
	\frac{\nabla_\mu X\,\nabla^\mu X}{4X} \;\longrightarrow\; 1 \,.
\end{equation}
Assuming the metric functions in \eqref{metric_ansatz}, the above condition reduces to, 
\begin{equation}
	F_1-F_2\big|_{\theta=0} = 0, 
    \label{eq:conic_sing}
\end{equation}
ensuring, therefore
\begin{eqnarray}
	\label{regular}
	\lim_{\theta \to 0}\frac{g_{\varphi\varphi}}{g_{\theta\theta}}
	= \theta^{2} + \dots,
\end{eqnarray}
so that the metric locally approaches the flat metric in spherical coordinates. 
In addition, the metric functions should satisfy Neumann boundary conditions on the  symmetry axis,
\begin{equation} \label{eq:derivative_pole_metric}
	\partial_\theta F_i\big|_{\theta=0} = \partial_\theta W\big|_{\theta=0} = 0\, ,
\end{equation}
which follow from a study of an approximate form of the solution. The boundary condition for the scalar amplitude on the axis distinguishes the two families,
\begin{equation}\label{eq:derivative_pole_scalar}
\phi\big|_{\theta=0} =0~~{\rm for~RBSs},\qquad \partial_\theta\phi\big|_{\theta=0} =0~~{\rm for~DBSs}.
\end{equation}
These conditions follow from the local expansion analysis mentioned above: $\phi\sim(r\sin\theta)^{|m|}$ for RBSs, and $\phi\sim r\cos\theta$ for DBSs. 

\medskip
Finally, since we assume the solutions to have an energy density that is symmetric under the equatorial reflection $\theta\to\pi-\theta$, the metric functions should satisfy Neumann boundary conditions on the equatorial plane,
\begin{eqnarray}
&&\label{bc_eq}
\partial_\theta F_i\big|_{\theta=\pi/2} = \partial_\theta W\big|_{\theta=\pi/2} =0.
\end{eqnarray}
For the scalar field amplitude, both parities under the equatorial reflection are compatible with a symmetric energy density. The corresponding boundary condition again distinguishes the two families of solutions: the scalar field is even for the RBSs\footnote{We mention that RBSs with a scalar field that is 
odd under reflection across the equatorial plane do also exist  \cite{Kleihaus:2005me}.} and odd for the DBSs,
\begin{eqnarray}
\label{bc_eq1}
    \partial_\theta \phi\big|_{\theta=\pi/2} = 0~~~{\rm for~RBSs},\qquad \phi\big|_{\theta=\pi/2} = 0~~~{\rm for~DBSs}.
\end{eqnarray}


In all solvers, we impose the boundary conditions given by Eqs.~\eqref{bc_inf} and \eqref{bc_inf1} at $r\to\infty$, Eq.~\eqref{r=0} at $r=0$, Eqs.~\eqref{bc_eq} and ~\eqref{bc_eq1} at $\theta=\pi/2$, and Eq.~\eqref{eq:derivative_pole_scalar} for the scalar field at $\theta=0$. For the metric functions on the symmetry axis, however, the choice of boundary conditions is not unique and we have considered the following two possibilities:
\begin{align*}
    {\rm I}:&\quad \partial_\theta F_0\big|_{\theta=0} =\partial_\theta F_1\big|_{\theta=0} =\partial_\theta F_2\big|_{\theta=0} = \partial_\theta W\big|_{\theta=0} = 0, \\
    {\rm II}:&\quad \partial_\theta F_0\big|_{\theta=0}=\partial_\theta F_2\big|_{\theta=0}=\partial_\theta W\big|_{\theta=0}=0,\;\; F_1\big|_{\theta=0}=F_2\big|_{\theta=0}.
\end{align*}
Choice I is the same as in Eq.~\eqref{eq:derivative_pole_metric}, namely Neumann boundary conditions for all metric functions, whereas in choice II, the condition $\partial_\theta F_1\big|_{\theta=0}=0$ is replaced by the no-conical singularity condition \eqref{eq:conic_sing}. 

A local expansion analysis near $\theta=0$ reveals that under the requirement of having a regular solution, both the Neumann condition and the absence of conical singularity should hold simultaneously by virtue of the field equations. However, in practice, depending on the numerical framework, one of these two choices may lead to better convergence properties. We therefore adopt the same set of boundary conditions for all functions in all solvers, except for the treatment of $F_1$ on the symmetry axis, see Sec.~\ref{sec:specs}. As a result, the three numerical approaches do not solve, \textit{a priori}, exactly the same boundary value problem: the only difference being the condition imposed at $\theta=0$. Nevertheless, for each solver, we verify \textit{a posteriori} that the remaining condition, not explicitly imposed during the computations (either $\partial_\theta F_1\big|_{\theta=0}=0$ or $F_1-F_2\big|_{\theta=0}=0$), is satisfied by the converged solutions within consistent numerical accuracy\footnote{In practice, we verify that the corresponding residual is comparable to the numerical errors estimated from the different code-independent diagnostics defined in Sec.~\ref{sec:err_estim}.}.

%
\subsection{Quantities of interest and units}

The ADM mass $M$ and the angular momentum $J$ are read from 
the asymptotic sub-leading behaviour of the metric functions, see Eq.~\eqref{AF-expansion},
\begin{eqnarray}
\label{asym}
g_{tt} =-e^{2F_0}+e^{2F_2}W^2r^2 \sin^2 \theta
=-1+\frac{2GM}{r}+\dots,~~g_{\varphi t}=-e^{2F_2}W r^2 \sin^2 \theta=-\frac{2GJ}{r}\sin^2\theta+\nonumber \dots. \\
\end{eqnarray}  
Alternatively, the mass $M$ can also be computed as an integral over a spacelike slice $\Sigma$,
\begin{eqnarray}
\label{Mpsi}
M= \int_{\Sigma} dS_a (T\xi^a-2T_{b}^a \xi^b)=
 4\pi \int_{0}^\infty dr \int_0^\pi d\theta~r^2\sin \theta ~e^{F_0+2F_1+F_2}
 \left(
 -\mu^2+2 e^{-2F_0} {\omega(\omega-mW)} 
 \right)\phi^2.  ~~
\end{eqnarray}

As for the Noether charge $Q$,
a straightforward computation shows that the angular momentum density and the Noether charge density are proportional,
$T_\varphi^t=m j^t$.
Then, using the Tolman expressions for the total 
angular momentum
and the expression  (\ref{Q})
for the Noether charge,
the following relation holds
\cite{Schunck:1996he},
\begin{eqnarray} 
\label{JQrel}
J=\int_\Sigma d^3 x  \sqrt{-g}\,T_\varphi^t = m Q,
\end{eqnarray}
with the explicit form
\begin{eqnarray}
\label{Q-int}
Q=4\pi \int_{0}^\infty dr \int_0^\pi d\theta  
~r^2\sin \theta ~e^{-F_0+2F_1+F_2}   {(\omega-mW)} \phi^2.
\end{eqnarray}
 
 As usual when dealing with gravitating massive scalar fields, 
the numerical integration is performed 
with dimensionless variables
introduced  
by using
natural units set by $\mu$ and $G$,
\begin{eqnarray}
r\to \mu\,r,~~
\phi \to \phi M_{Pl}/\sqrt{4\pi},~~
\omega \to \omega/\mu, 
 \end{eqnarray}
 where $M_{Pl}^2=G^{-1}$ is the Planck mass.
 As a result, the dependence on both $G$ and $\mu$
 disappears from the equations.
 Also, the global charges and all other quantities of interest are 
 expressed in units set by $\mu$ and $G$ (note that,
 in order to simplify the output,
  we set $G=1$ in what follows).


\section{Methods and software}

\label{sec:methods_software}


\subsection{Finite differences: { the \texttt{FIDISOL/CADSOL} solver}}\label{sec:fidisol_cadsol}

We start by presenting a concise review of finite-difference methods for stationary, second-order, nonlinear, coupled elliptic PDEs in two spatial dimensions using the \texttt{CADSOL} (Cartesian Arbitrary Domain SOLver)
 package written in Fortran \cite{SCHONAUER2001473,schmauder1992cadsol,SCHONAUER1989279}. The code was developed from the late 1980s through the early 2000's by a group from  Karlsruhe University in Germany, being an extension of the \texttt{FIDISOL} (FInite DIfference SOLver) package produced by the same team. 
\texttt{CADSOL} focuses on 2D/3D elliptic/parabolic systems posed on arbitrary physical domains via body-oriented (index-rectangular) grids; here the index grid and finite-difference order are user-prescribed. Spatial derivatives are approximated by finite-difference formulas generated from local interpolation polynomials, nonlinear systems are solved by Newton-Raphson, and the resulting linear systems are handled by the polyalgorithmic \texttt{LINSOL} package. The solver computes \textit{a posteriori} discretization-error estimates based on differences between the active formulae of consistency order \texttt{PD} and control formulae of order \texttt{PD}+2.

To turn \eqref{eq:cadsol-operator} into a set of algebraic equations, the fields $\mathcal{F}$ are represented by their values, $U_{i,j}$, on a logically rectangular index grid $(i,j)$, with $i=1{:}N_X$ and $j=1{:}N_Y$ (non-equidistant spacings are allowed). The user provides coordinate arrays $(X(i,j),\,Y(i,j))$ that map each index node to a physical point of $\Omega$. Boundary conditions are prescribed through boundary operators $G^{(s)}=0$, where $s=1,\cdots,4$ labels the four sides of the rectangular index grid. In \texttt{CADSOL}, these correspond respectively to the boundaries $i=1$, $i=N_X$, $j=1$, and $j=N_Y$. Each $G^{(s)}$ may encode Dirichlet, Neumann, or Robin conditions (and nonlinear variants).

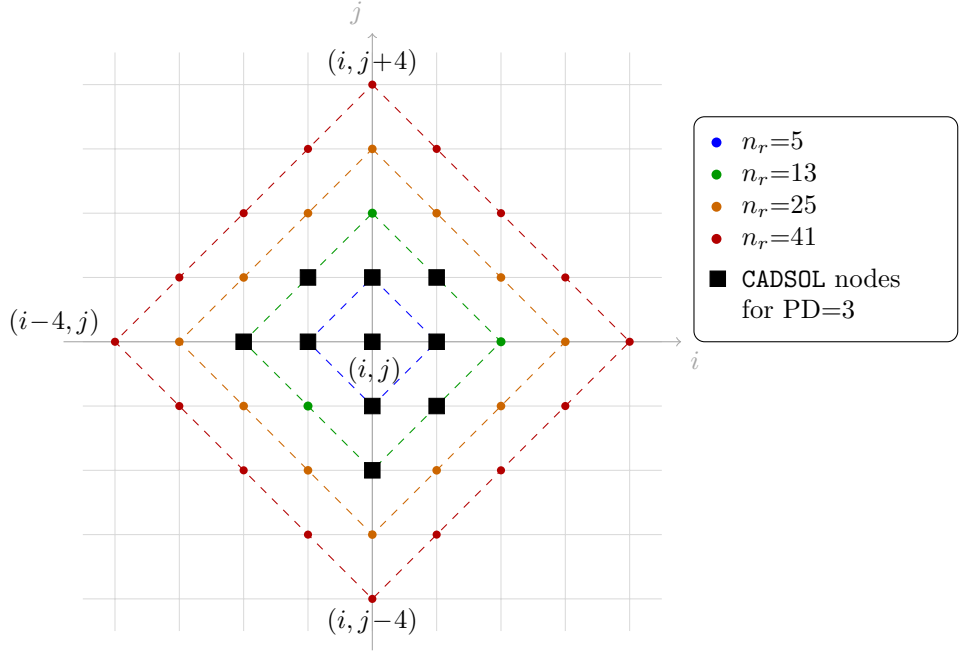
\begin{figure}[ht]
\centering
\begin{tikzpicture}[scale=0.85]
	\draw[step=1.cm,very thin,gray!30] (-4.5,-4.5) grid (4.5,4.5);
	\draw[->,gray!70] (-4.8,0) -- (4.8,0) node[below right] {$i$};
	\draw[->,gray!70] (0,-4.8) -- (0,4.8) node[above left] {$j$};
	
	\fill[black] (0,0) circle (2.6pt) node[below right]{};
	
	\foreach \x/\y in {1/0,-1/0,0/1,0/-1} \fill[blue] (\x,\y) circle (2.2pt);
	\foreach \x/\y in {2/0,-2/0,0/2,0/-2,1/1,1/-1,-1/1,-1/-1}
	\fill[green!60!black] (\x,\y) circle (2pt);
	\foreach \x/\y in {3/0,-3/0,0/3,0/-3,2/1,2/-1,-2/1,-2/-1,1/2,1/-2,-1/2,-1/-2}
	\fill[orange!80!black] (\x,\y) circle (1.9pt);
	\foreach \x/\y in {4/0,-4/0,0/4,0/-4,3/1,3/-1,-3/1,-3/-1,2/2,-2/2,2/-2,-2/-2,1/3,1/-3,-1/3,-1/-3}
	\fill[red!70!black] (\x,\y) circle (1.8pt);
	
	\foreach \k/\col in {1/blue,2/green!60!black,3/orange!80!black,4/red!70!black}{
		\draw[dashed,\col] (0,\k) -- (\k,0) -- (0,-\k) -- (-\k,0) -- cycle;
	}
	
	\node[anchor=north] at (0,-4) {$(i,j\!-\!4)$};
	\node[anchor=west]  at (-5.8,0.3)  {$(i\!-\!4,j)$};
	\node[anchor=south east] at (0.6,-0.8) {$(i,j)$};
	\node[anchor=south] at (0,4) {$(i,j\!+\!4)$};
	
	\begin{scope}[xshift=5.0cm,yshift=0.cm]
		\draw[fill=white,rounded corners,draw=black] (0,0) rectangle (4.1,3.5);
		\fill[blue] (0.35,3.1) circle (2.2pt); \node[right] at (0.6,3.1) { $n_r{=}5$};
		\fill[green!60!black] (0.35,2.6) circle (2.2pt); \node[right] at (0.6,2.6) {$n_r{=}13$};
		\fill[orange!80!black] (0.35,2.1) circle (2.2pt); \node[right] at (0.6,2.1) {$n_r{=}25$};
		\fill[red!70!black] (0.35,1.6) circle (2.2pt); \node[right] at (0.6,1.6) {$n_r{=}41$};
		\draw[black,fill=black] (0.25,0.86) rectangle (0.49,1.1);
		\node[right,align=left] at (0.6,0.75) {\texttt{CADSOL} nodes \\for PD=3};
		
	\end{scope}
	
	\foreach \x/\y in {-2/0,-1/0,0/0,1/0, -1/1,0/1,1/1, 0/-1,1/-1, 0/-2}{
		\draw[black,fill=black] (\x-0.12,\y-0.12) rectangle (\x+0.12,\y+0.12);
	}
	
\end{tikzpicture}

\caption{\texttt{CADSOL} difference star of order 3 (adapted from \cite{schmauder1992cadsol}).%
	\label{fig:cadsol-stencil}}

\end{figure}

The user selects the consistency order \texttt{PD}=n (polynomial degree $n$). Around each node in the grid $k=(i,j)$ with physical location $(X_k,Y_k)$, \texttt{CADSOL} automatically selects a set of $m_n=\frac{(n+1)(n+2)}{2}$ nearby nodes, called the difference star $S^{(n)}_{i,j}$, with $k$ as its central point (see Fig. \ref{fig:cadsol-stencil}). Using the solution values on these points, it constructs a local 2-D interpolation
polynomial of order n, $P_{n,k}(X,Y)$, defined through
\begin{equation}
    P_{n,k}(X,Y)=a_{1,k}+a_{2,k} X+a_{3,k} Y+a_{4,k} X^2+a_{5,k} Y^2+a_{6,k} X Y+\cdots
\end{equation}

Let $\tilde{k}=(\tilde{i},\tilde{j})$ be a nearby point to $k$ lying in its difference star, the polynomials are normalized such that $P_{n,k}(X_{\tilde{k}},Y_{\tilde{k}})=\delta_{k \tilde{k}}$. 
All spatial derivatives required by $\mathcal{E}$ at $(i,j)$ are then obtained by differentiating this same polynomial and evaluating the result at the star's center, so every finite-difference formula at that node comes from the same local approximation. Higher-order stars are built as extensions of lower-order ones, which \texttt{CADSOL} uses to estimate the discretization error by comparing formulas of different order. Specifically, let $U_l=U(X_l,Y_l)$ denote the solution values at the $m_n$ star nodes, $l=1,\dots,m_n$. The degree-$n$
\emph{local interpolant} $U_d$ and the difference formulas for its derivatives are given by
\begin{equation}
U_d(X,Y)=\sum_{l=1}^{m_n} U_l\,P_{n,l}(X,Y)\,,\qquad U_{d,XY}(X,Y)
= \sum_{l=1}^{m_n} U_l
\,\bigl(\partial_{XY} P_{n,l}\bigr)(X,Y)\,.
\end{equation}

Here the second expression is shown for the mixed derivative as an
example; the formulas for \(U_{d,X}\), \(U_{d,Y}\), \(U_{d,XX}\), and
\(U_{d,YY}\) are obtained analogously by differentiating the same local
interpolation polynomials with respect to the appropriate variables. Notice also that, due to the normalization of the polynomials, we have $U_d(X_k,Y_k)=U_k$. For each $(i,j)$, the star $S^{(n)}_{i,j}$ is chosen so that the local interpolation
system is regular (the nestedness of the stars is needed by the error estimator below). On rectangular grids
the interior stars are generated from a triangular pattern and extended
systematically with increasing order; near physical boundaries the pattern is
shifted inward to remain inside the domain while
preserving linear independence. The admissible neighbors around a center are
confined to a rhomb in index space containing at most,
\begin{equation}
n_r \;=\; 2\,n(n+1)+1 \qquad \text{(code parameter \texttt{NCC})}
\end{equation}
candidate points; the actual star uses exactly $m_n$ of them.

With the operators built from the degree-\texttt{PD} star, we define the nodal residual evaluated at the node $k=(i,j)$,
\begin{equation}
R_{k}(U)\;=\;\mathcal{E}\Bigl(X_k,Y_k;\ U_{d},\ U_{d, X},\ U_{d, Y},\ U_{d, XX},\ U_{d, XY},\ U_{d, YY}\Bigr),
\end{equation}

 The discrete problem therefore seeks the grid values of $U$
such that $R_k(U)=0$ at every node $k$. Collecting all nodes and all $N_K$ solution components, i.e. all
unknown fields, gives the global residual vector $R(U)$. In practice we solve the nonlinear system by damped Newton,
\begin{equation}\label{JAC}
J\!\bigl(U^{(\nu)}\bigr)\,\Delta U^{(\nu)}=-R\!\bigl(U^{(\nu)}\bigr),\qquad
U^{(\nu+1)}=U^{(\nu)}+\alpha\,\Delta U^{(\nu)},\quad \alpha\in(0,1].
\end{equation}

The Jacobian, $J$,  is assembled analytically from the user-supplied derivatives of $\mathcal{E}$ and $G^{(s)}$; $\alpha=1$ is typical and reduced only for robustness. 
At each Newton step, that is, when solving for the correction $\Delta U^{(\nu)}$ in the Jacobian system \eqref{JAC} above, \texttt{CADSOL} uses the \textsc{LINSOL} package, a polyalgorithm of Conjugate-Gradient and other Krylov-subspace methods with automatic fallback.

\texttt{CADSOL} estimates the discretization error together with the solution. For each derivative, it compares the finite-difference formula of order \texttt{PD} with the corresponding formula of order \texttt{PD}+2. Through the appropriate Jacobian entries, these estimates contribute to the stopping criterion of the Newton iteration, which terminates when the residual is sufficiently small compared with both the estimated discretization error and the tolerance \texttt{TOL}. After convergence, they are used on the right-hand side of a further linear solve with the Newton matrix to obtain the \textit{a posteriori} error estimate. \texttt{CADSOL} also provides the estimate at every grid point as a relative error for each solution component, normalized by the maximum magnitude of that component over the grid.

\subsection{Finite elements:  { the \texttt{FreeFem} solver}}

\label{fem}
We present a concise overview of the finite element method (FEM) for solving stationary, second-order, nonlinear, coupled elliptic PDEs in two spatial dimensions. In this approach, the original continuous problem (of general form~\eqref{eq:cadsol-operator}) 
is discretized by restricting it to a finite-dimensional space of functions -- the \textit{finite element space} -- spanned by basis functions defined using a mesh covering the spatial domain. The unknown fields are expanded in this basis, converting the PDEs into a system of algebraic equations for the expansion coefficients. Unlike finite-difference schemes (and many others), the FEM does not rely on the usual \textit{strong} form of PDEs, but rather on their \textit{weak} (or \textit{variational}) formulation, expressed as an integral equation over the spatial domain. 

Our implementation is based on the open-source \texttt{FreeFem} software written in C++ \cite{MR3043640}. Originally developed in the late 1980s by Frédéric Hecht and collaborators at Pierre and Marie Curie University in Paris, \texttt{FreeFem} has evolved into a versatile computational framework for research and engineering applications. It supports general unstructured triangular (2D) and tetrahedral (3D) meshes\footnote{Recent developments also enables solving one- and two-dimensional problems posed on arbitrary curved manifold embedded in $\mathbb{R}^3$.} generated by the internal \texttt{BAMG} mesher, whose flexible meshing capabilities supplemented by \texttt{FreeFem}'s built-in interpolation between meshes enable efficient adaptive refinement strategies, allowing the mesh to be tailored to specific field behaviors and thereby improving numerical accuracy. \texttt{FreeFem}'s high-level scripting language enables the specification of weak forms of PDEs using a syntax that closely mirrors the underlying mathematics. A variety of finite element spaces consisting of piecewise polynomial functions, several linear solvers for matrix inversion, and a number of optimization routines are provided natively. 


We now outline the procedure for deriving a weak form of the field equations relevant to this work. 
Both the \texttt{FreeFem} and \texttt{CADSOL} implementations
employ a  compactified radial coordinate $x$  defined as 
\begin{equation}
\label{comp}
    x=\frac{r}{r+c}, 
\end{equation}
which maps the semi-infinite interval $r\in[0,\infty)$ onto the finite segment $x\in[0,1]$. Here, $c$ is the \textit{compactification} parameter typically set to one, but adjustable to improve numerical accuracy in specific regimes.
Then, all equations \eqref{eq-phi}-\eqref{eq-W} can be cast in the form,
\begin{equation}
    \mathcal{E}_{\cal F}\equiv e^{-2F_1}\left[\frac{(1-x)^4}{c^2}{\cal F}_{,xx}+\left(\frac{1-x}{cx}\right)^2{\cal F}_{,\theta\theta}\right]+\dots=0,
\end{equation}
where the dots denote terms with derivatives of order lower than two. We include the overall factor $e^{-2F_1}$ because it coincides with that appearing in the principal part of the Laplace-Beltrami operator $\Delta {\cal F}\equiv\partial_\mu(\sqrt{-g}\,g^{\mu\nu}\partial_\nu{\cal F})/\sqrt{-g}$ for the metric in~\eqref{metric_ansatz}. For each field variable $\cal F$, we introduce a \textit{test function} $v_{\cal F}=\{v_\phi,v_{F_{0}},v_{F_{1}},v_{F_{2}},v_W\}$ defined over the computational domain $\Omega$. Each $v_{\cal F}$ is arbitrary within $\Omega$, but must vanish on boundaries where the corresponding field variable ${\cal F}$ satisfies Dirichlet conditions (for example, $v_{\cal F}=0$ at spatial infinity, $x=1$, see Eqs.~\eqref{bc_inf} and \eqref{bc_inf1}). Multiplying each equation $\mathcal{E}_{\cal F}$ by $v_{\cal F}$, integrating over $\Omega$, and integrating by parts terms with second-derivatives yields,
\begin{equation}
\label{weak_form_1}
     -\int_\Omega{dx\,d\theta\sqrt{-g}\,v_{\cal F}\,{\cal E}_{\cal F}}=\int_\Omega{dx\,d\theta\sqrt{-g}\left(e^{-2F_1}\nabla{\cal F}\cdot\nabla v_{\cal F}+\dots\right)}=0,
\end{equation}
where we have introduced the notation $\nabla S\cdot\nabla T\equiv \frac{(1-x)^4}{c^2}S_{,x}T_{,x}+\left(\frac{1-x}{cx}\right)^2S_{,\theta}T_{,\theta}$. Notice the absence of boundary terms, which vanish by virtue of the boundary conditions\footnote{For more general boundary conditions, such as Robin or inhomogeneous Neumann types, the boundary terms may not vanish. One might say that the boundary conditions within the weak formulation are naturally implemented by boundary integrals.}. The weak formulation of the original problem is then obtained by summing all individual integral equations of the form \eqref{weak_form_1} and requiring the resulting expression to vanish for arbitrary test functions $v_\phi,\dots,v_W$. As a result, the field equations are implemented in the code as a single \textit{master} integral equation. This is facilitated by \texttt{FreeFem}'s built-in routines for integration and differentiation which exploit the finite element discretization presented below.

\begin{figure}
\centering
  \includegraphics[scale=0.33]{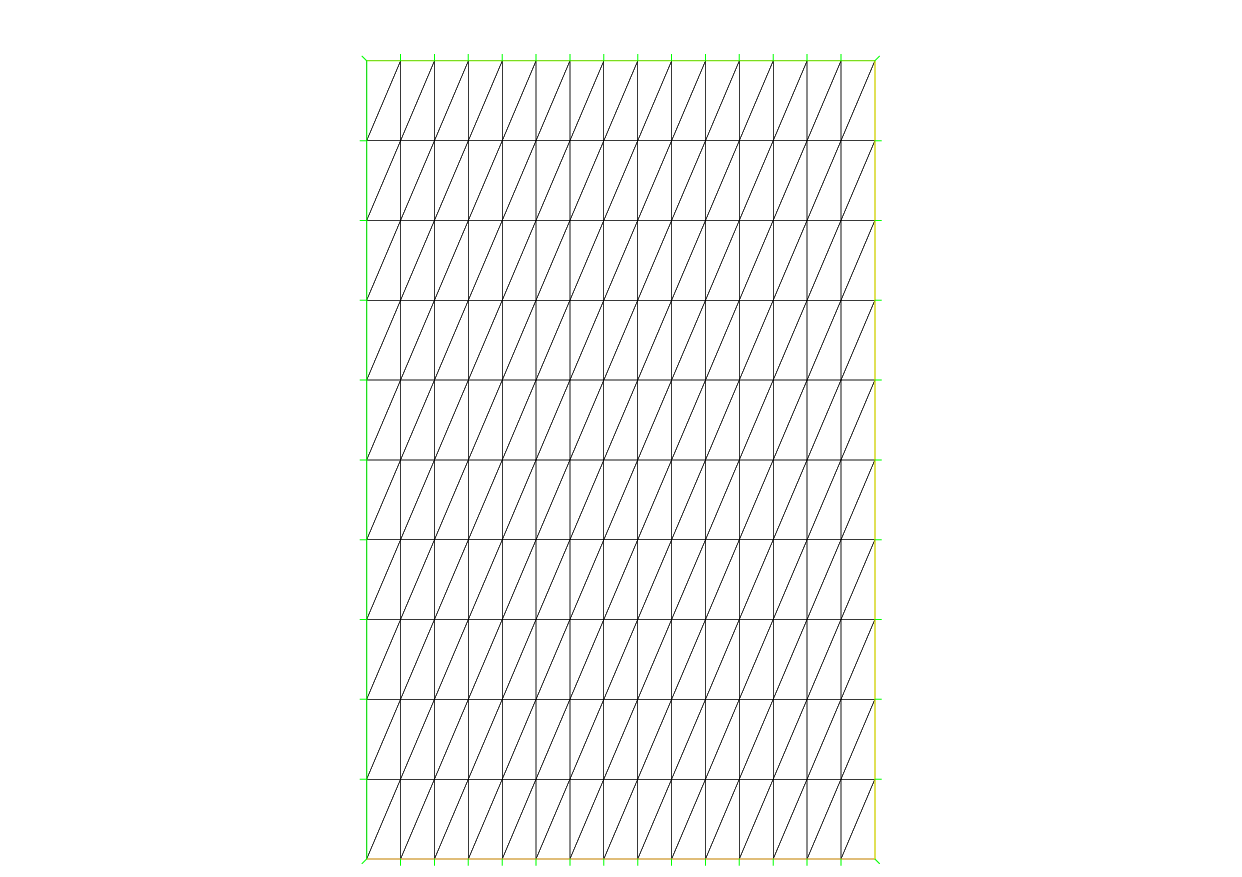}\hspace{-2cm}
  \includegraphics[scale=0.33]{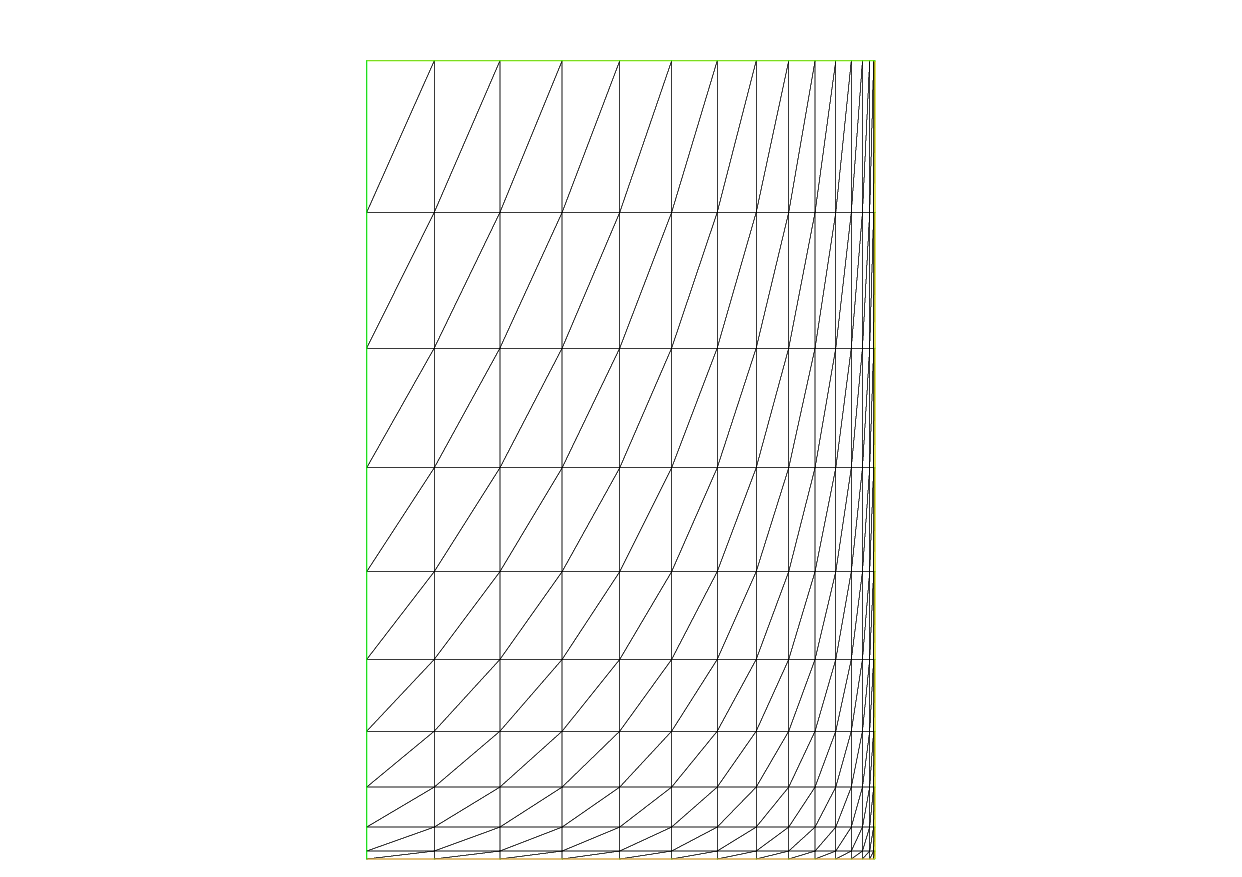}\hspace{-2cm}
  \includegraphics[scale=0.33]{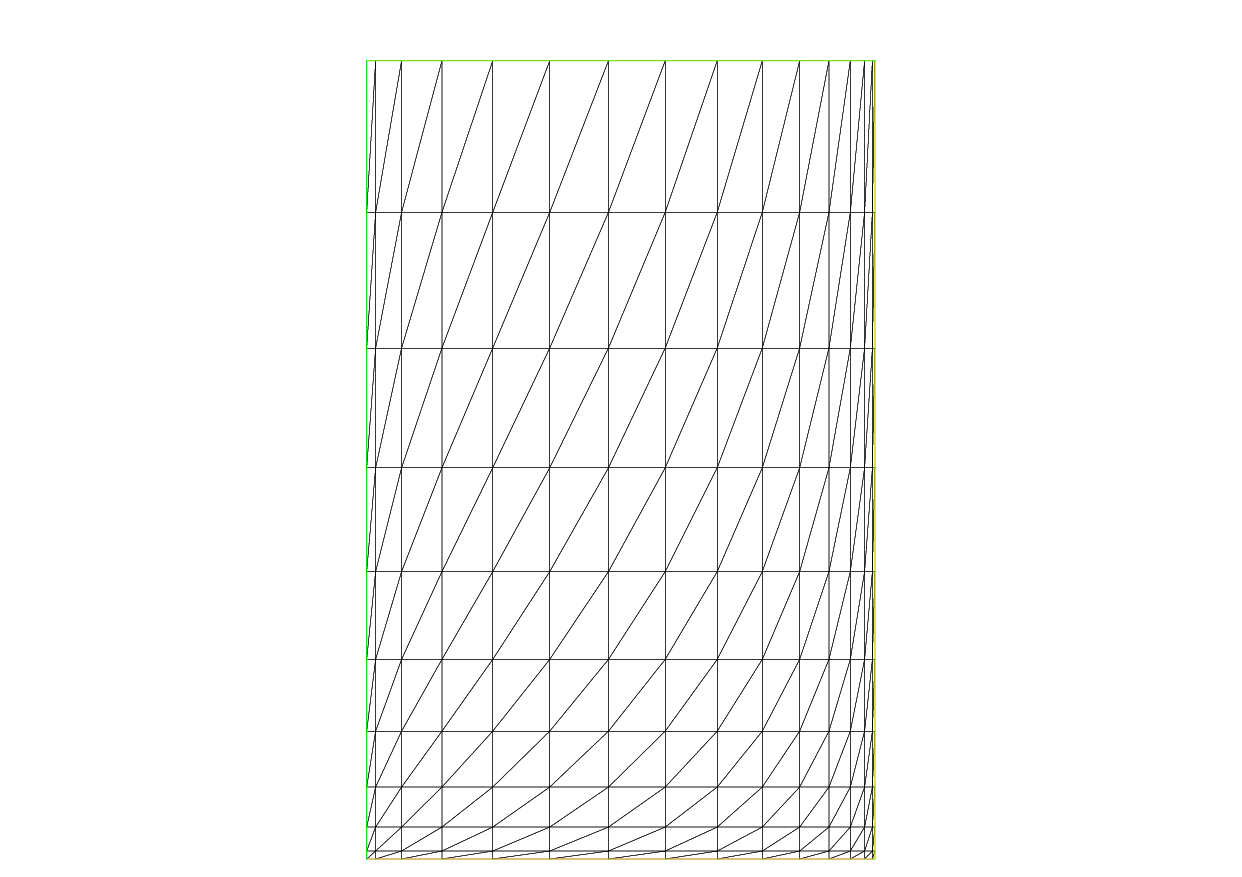}
  \caption{Examples of \texttt{FreeFem}-generated meshes for a rectangular domain $\Omega$, illustrating uniform (left) and nonuniform (center and right) cell distributions.}
  \label{meshes}
\end{figure}

For the purpose of this work, the computational domain $\Omega$ is a rectangular region in the $(x,\theta)$ coordinate plane. Using a built-in routine, $\Omega$ is discretized by a mesh with triangular cells (see Fig.~\ref{meshes}) characterized by two resolution parameters, $N_x$ and $N_\theta$, denoting the number of edges along the $x$ and $\theta$ directions, respectively. Discretization is then achieved by requiring the field variables $\cal F$ and the test functions $v_{\cal F}$ to belong to a finite element space $V_h$, defined as the space of continuous functions whose restriction to each mesh cell is a polynomial of degree $k$ in the coordinates $x$ and $\theta$. Introducing a basis $\mathfrak{B}=(\phi_1,\dots,\phi_M)$ of $V_h$, with $M\equiv\text{dim}(V_h)$, $\cal F$ and $v_{\cal F}$ are expanded as,
\begin{equation}
\label{decomp_fe}
    {\cal F}(x,\theta)=\sum_{i=1}^{M}{{\cal F}_i\,\phi_i(x,\theta)},\quad\quad v_{\cal F}(x,\theta)=\sum_{i=1}^{M}{v_i\,\phi_i}(x,\theta),
\end{equation}
where ${\cal F}_i,\,v_i\in\mathbb{R}$ are the expansion coefficients in the \textit{finite element basis} $\mathfrak{B}$. \texttt{FreeFem} internally represents functions in terms of these coefficients. Substituting the expansion \eqref{decomp_fe} of the test function into Eq.~\eqref{weak_form_1} gives,
\begin{equation}
\label{non_lin_syst}
    \sum_{i=1}^M{v_i\,E_i({\cal F})}=0\quad\text{with}\quad E_i({\cal F})=\int_\Omega{dx\,d\theta\sqrt{-g}\left(e^{-2F_1}\nabla{\cal F}\cdot\nabla\phi_i+\dots\right)}.
\end{equation}
Since the above equation must hold for arbitrary $v_i$, the problem reduces to solving $E_i({\cal F})=0$, which defines a nonlinear algebraic system for the coefficients ${\cal F}_i$ in \eqref{decomp_fe}. In practice, we solve this system using the standard Newton-Raphson method. Starting from an initial guess for the expansion coefficients of the field variables, $U^{(0)}\equiv({\cal F}_1^{(0)},\dots,{\cal F}_M^{(0)})$, we compute the successive values $U^{(\nu)}$ via the recursive formula,
\begin{equation}
\label{eq:newt_step}
    U^{(\nu+1)}=U^{(\nu)}-J^{-1}\big(U^{(\nu)}\big)\,E\big(U^{(\nu)}\big),
\end{equation}
where $E$ denotes the vector with components $E_i$ defined in Eq.~\eqref{non_lin_syst}, and $J$ is the Jacobian matrix with entries $J_{ij}({\cal F})=\partial E_i/\partial{\cal F}_j$. We stop the iterations when the step size $\left\lVert U^{(\nu+1)}-U^{(\nu)}\right\rVert$ falls below a prescribed tolerance. 

\subsection{Spectral method: {  the \texttt{Kadath} solver}}
\label{spectral_Kadath}

\texttt{Kadath} is an open-source C++ library implementing multi-domain spectral methods developed at Paris-PSL Observatory by Philippe Grandcl\'ement~\cite{Grandclement:2009ju}. In the present work we employ the publicly available BS solver 
implemented by Grandcl\'ement in \cite{Grandclement:2014msa}, together with some adaptations specific to the configurations studied here. The solver uses a 2D axisymmetric ``polar’’ setup with coordinates $(r,\theta)$, where the spatial domain is decomposed into a set of touching but non-overlapping spherical subdomains in the radial direction. 
The outermost domain is compactified so as to extend up to spatial infinity, while the inner domains cover the near-origin and intermediate regions with spectral resolution adapted to the field gradients. As in the other solvers presented in this work, compactified coordinates are employed to impose asymptotic flatness, although in \texttt{Kadath} this compactification is restricted to the external domain only. 
For the stationary axisymmetric configurations considered here, the dependence on the azimuthal angle $\varphi$ is absent. This follows from the use of the \texttt{Space\_polar} class in the BS solver, which exploits axisymmetry to reduce the full three-dimensional \texttt{Space\_spheric} geometry, also available in the public library, to an effective two-dimensional spectral decomposition. We also note that \texttt{Kadath} provides several additional geometries based on different coordinate systems.

Within each domain, the unknown fields are represented through truncated spectral expansions,
\begin{equation}\label{eq:spectral_exp}
{\cal F}(r,\theta)
=
\sum_{n,\ell}
c_{n\ell}\,
R_n(r)\,
\Theta_\ell(\theta)\,
\end{equation}
where $R_n(r)$ are Chebyshev polynomials (or combinations thereof adapted to the domain topology and boundary conditions), while $\Theta_\ell(\theta)$ 
are trigonometric basis functions.
Regularity conditions at the origin and on the symmetry axis are enforced through suitable parity choices in the spectral basis. In particular, even Chebyshev polynomials are used on the numerical domain that contains the origin, cosine angular modes are employed for fields that remain finite on the axis, whereas sine expansions are used for quantities that vanish there. Continuity of the fields and of their normal derivatives across domain interfaces is imposed automatically by the library through matching equations added to the global nonlinear system.

The equations implemented in \texttt{Kadath} are not directly those presented in Section~\ref{eom}, but rather the reformulated system described below in Section~\ref{spec_kadath}, specifically adapted to the spectral decomposition employed by the library. The unknowns are the spectral coefficients $c_{n\ell}$, and the equations are solved using a spectral Galerkin method in coefficient space, with differential operators evaluated through spectral differentiation matrices. The spectral basis determines a set of Gauss–Lobatto collocation points, which provide the interpolation nodes between the spectral and physical representations of the fields. Unlike finite-difference or finite-element methods, these points are not chosen independently by the user but are fixed once the spectral basis and truncation order are specified. The resulting discretization yields a coupled nonlinear algebraic system for the spectral coefficients. As in \cite{Grandclement:2014msa}, this system is solved through a Newton–Raphson iteration in which all fields are updated simultaneously. The Jacobian matrix is assembled automatically from the symbolic form of the equations provided by the user, and the corresponding linear system is solved using \texttt{LAPACK} routines. Iterations are stopped once the residual falls below a prescribed tolerance or reaches the round-off level. 

Owing to the spectral discretization, the method exhibits exponential convergence for sufficiently smooth solutions, as expected for stationary BS configurations; explicit convergence tests are presented in Section~\ref{sec:err_estim}. Depending on the particular BS configuration under consideration, different decompositions into spherical subdomains are employed, with domain boundaries placed at radial locations specified below. Across all domains we use a fixed number of radial and angular spectral coefficients. More precisely, in the expansion \eqref{eq:spectral_exp} we truncate the spectral series according to $n\in[0,N_r-1]$ and $\ell\in[0,N_\theta-1]$.

Boundary conditions corresponding to regularity at $r=0$, axisymmetry, equatorial symmetry, and asymptotic flatness are imposed through the choice of spectral basis.
In order to prevent convergence to the trivial vacuum solution $\phi=0$, the first configuration of a given BS sequence is obtained by fixing the value of $F_0$ at the origin to a prescribed value close to zero (the Minkowski limit) while treating the frequency $\omega$ as an additional unknown. This drives the solver toward a weakly relativistic configuration in the Newtonian regime, $\omega/\mu\to1$, which is then used as the starting point for continuation along the family of solutions. Following the strategy implemented in the public \texttt{Kadath} solver \cite{Grandclement:2014msa}, convergence is further facilitated by first solving a partially Newtonian system for $\phi$ and $F_0$ (using the metric coefficient $F_0$ as the non-relativistic gravitational potential), taking $F_1=F_2=W=0$, and subsequently using the resulting configuration as an initial seed for the fully relativistic problem. 
Our modifications to the public implementation are limited mainly to changes in notation, sequence generation strategy, location of the domains as specified in Section~\ref{spec_kadath}, parity assignments for $\phi$ (in the case of DBSs), and output conventions, while preserving the original multi-domain spectral infrastructure and nonlinear solution strategy of \texttt{Kadath} \cite{Grandclement:2009ju}.

\section{Properties of the solutions sequences} \label{sec:gen_properties}


\begin{figure}[ht]
\includegraphics[width=0.48\linewidth]{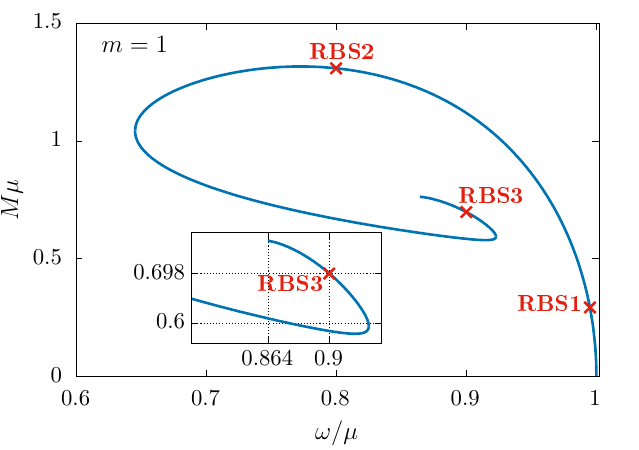}
\qquad{}
\includegraphics[width=0.48\linewidth]{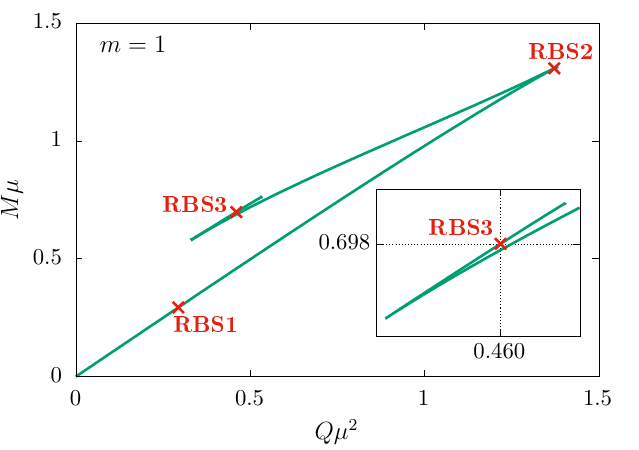}
\caption{\small 
{\it Left:}
The mass-frequency diagram is shown for the family of RBSs with azimuthal index $m=1$.
{\it Right:}
The mass-charge diagram for the same family of solutions. The red marks indicate three reference solutions \{RBS1, RBS2, RBS3\} for which the different solvers are compared.
}
\label{fig_RBS}
\end{figure}
%
%
\begin{figure}[ht]
\includegraphics[width=0.48\linewidth]{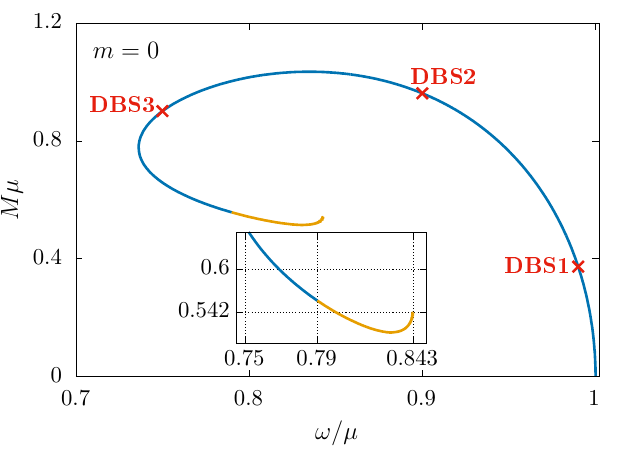}
\qquad{}
\includegraphics[width=0.48\linewidth]{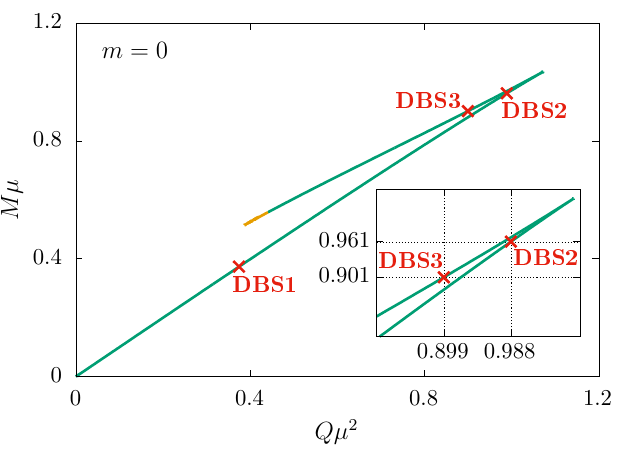}
\caption{\small 
{\it Left:}
The mass-frequency diagram is shown for the family of static DBSs.
{\it Right:}
The mass-charge diagram for the same family of solutions. The red marks indicate three reference solutions  \{DBS1, DBS2, DBS3\} for which the different solvers are compared.
The orange portions of the curves correspond to the part of the diagrams obtained for the first time in this work, using the \texttt{FreeFem} solver.
}
\label{fig_DBS}
\end{figure}

Before presenting the systematic comparison of the three solvers, we first review the general properties of two families of BS solutions considered in this work. 

\medskip

\begin{figure}[ht]
\includegraphics[width=0.48\linewidth]{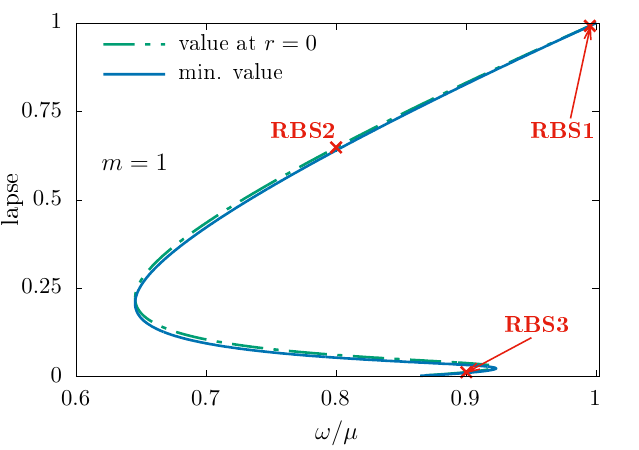}
\qquad{}
\includegraphics[width=0.48\linewidth]{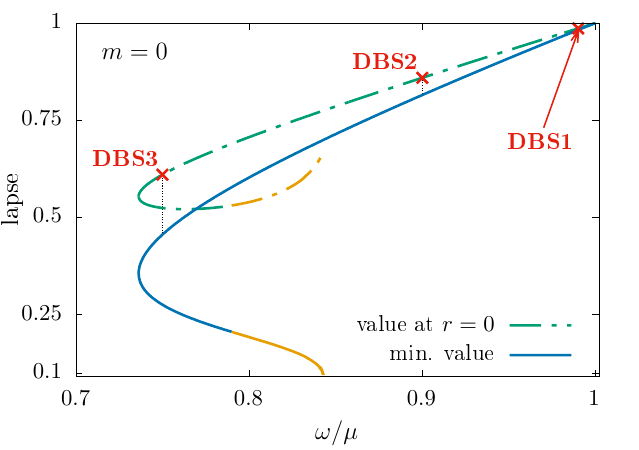}
\caption{\small 
The minimum value of the lapse $e^{F_0}$ is shown against the frequency for the families of RBSs (left) and DBSs (right); the value at the origin, $r=0$, is also shown for comparison. The six reference solutions are marked in red, while the newly extended portion of the second branch for DBSs is highlighted in orange. 
}
\label{fig_lapse}
\end{figure}

For both RBSs and DBSs
the solutions emerge from the vacuum at $\omega=\mu$ and exist for $\omega<\mu$;
decreasing the frequency,
 the mass increases  until a maximum value is reached,
 which is of the order of $1/\mu$.  Further decreasing $\omega$, one finds a minimal frequency  below which no BS solutions are found, see Figs. \ref{fig_RBS}, \ref{fig_DBS} (left panels).
The BS 
sequence further continues on the second branch, with $\omega$ increasing  while $M$ keeps decreasing, until a local minimum of the mass is reached.
Yet another branch emerges there,
with a backbending in $\omega$ and increasing in $M$.
While the numerics become increasingly challenging,
it is likely that,
similar to the spherically symmetric case \cite{Friedberg1987,Liebling:2012fv}, the $M(\omega)$ curve spirals towards a central region of the diagram. The curve $Q(\omega)$ (not shown here) exhibits a similar shape. As one moves toward the center of the spiral, the minimum value of the lapse $e^{F_0}$ decreases and approaches zero, see Fig.~\ref{fig_lapse}, indicating that the solutions near the spiral center are on the verge of forming an event horizon. While $M$ and $Q$ seem to remain finite throughout the sequences, it is likely that the limiting configurations at the center are singular (see, $e.g.$, Fig.~5 in Ref.~\cite{Gervalle2022} and discussion therein for the RBS case), which accounts for the increasing numerical difficulties encountered along the sequences. 

For RBSs, all three solvers successfully reach the third branch of solutions down to the value $\omega\simeq 0.86447$ (see the inset in left panel of Fig.~\ref{fig_RBS}), beyond which the quality of numerical solutions deteriorates. 
This third branch was previously constructed in Ref.~\cite{Herdeiro:2015gia}. 
For DBSs, the second backbending at the end of the second branch, occurring at $\omega\simeq 0.84284$, was reached exclusively with the \texttt{FreeFem} solver, the two other solvers failing to converge in this region. Higher branches in the static, dipolar case therefore appear to be the most difficult to construct numerically. The extended second branch, highlighted in orange in Figs.~\ref{fig_DBS},~\ref{fig_lapse}, has never been obtained before. 

\begin{figure}[ht]
\includegraphics[width=0.48\linewidth]{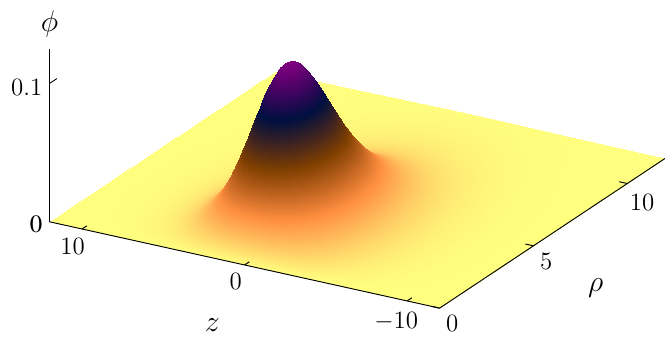}
\qquad{}
\includegraphics[width=0.35\linewidth]{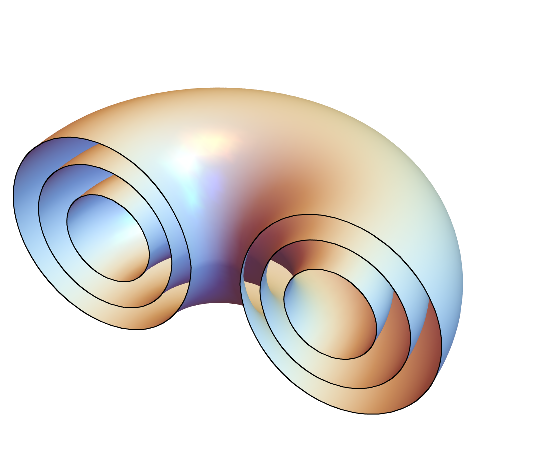}
\caption{\small 
 The scalar field profile and
surfaces  of constant energy density 
are shown
for the rotating solution RBS2. 
}
\label{spinningBS}
\end{figure}

\begin{figure}[ht]
\includegraphics[width=0.48\linewidth]{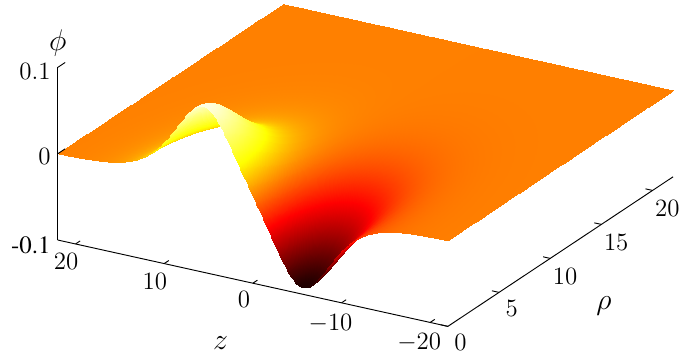}
 \hspace{2cm}
\includegraphics[width=0.20\linewidth]{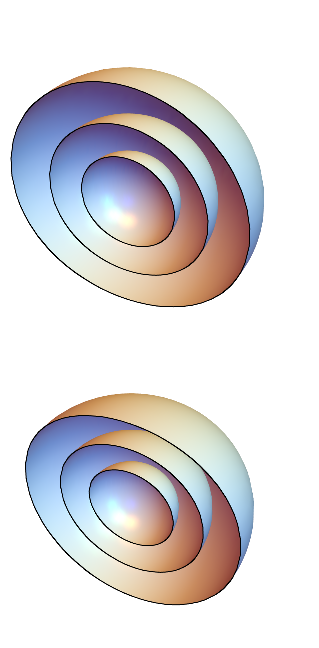}
\caption{\small  
 The scalar field profile and
surfaces  of constant energy density 
are shown
for the static solution DBS2. 
}
\label{DBS}
\end{figure}

The reason why the dipolar case is more challenging 
stems from the spatial localization of the fields at a finite distance from the origin along the symmetry axis (see Fig.~\ref{DBS}), where $\theta=0$. As can be seen from Eqs.~\eqref{eq-phi}-\eqref{eq-F0}, the effects of steep gradients around the axis are amplified by the factors $\cot\theta\simeq 1/\sin\theta$, a feature that appears to be better resolved using the \texttt{FreeFem}-based code. By contrast, RBSs are localized in the equatorial plane (see Fig.~\ref{spinningBS}), where $\theta=\pi/2$, which partially suppress steep gradient effects. At the same time, because of rotation, even highly relativistic RBSs remains localized at a small but \textit{finite} radial distance from the origin, which further alleviates these effects.

For a RBS, the typical surfaces of constant energy density 
 (as given by the $\tensor{T}{^t_t}$
 component of the energy-momentum tensor)
consist of coaxial tori 
located symmetrically $w.r.t.$ the equatorial plane, with the $z$-axis as their common axis of revolution (here $z=r \cos \theta$ and $\rho=r\sin\theta$).
For a DBS, a surface of constant energy density yields two spheroidal surfaces located along the
symmetry axis at $z=\pm z_0$ (where $z_0$ is location of maximum of $|\phi|$). In both cases, the maximum of the energy density coincides with that of the scalar field amplitude. 
These aspects are illustrated in Figs. \ref{spinningBS}, \ref{DBS}.

\medskip

In the remainder of this work, we shall compare results for three reference spinning solutions: RBS1, RBS2 and RBS3, with the input parameter $\omega/\mu=0.995$, $0.8$ (first branch) and $0.9$ (third branch), respectively.
Three reference dipolar configurations on the first branch are also considered: DBS1, DBS2, and DBS3, with $\omega/\mu=0.99$, $0.9$ and $0.75$, respectively.
The position of these six reference configurations is displayed on the diagrams in 
Figs.~\ref{fig_RBS},~\ref{fig_DBS},~\ref{fig_lapse}. These six reference configurations have been chosen to sample qualitatively distinct regimes of the two solution families, each placing different demands on the solvers. RBS1 and DBS1 lie in the weakly relativistic (Newtonian) regime, where the scalar field is dilute and spatially extended. RBS2 and DBS2 belong to the intermediate regime, close to the maximum mass along the fundamental branch, where the solutions are moderately compact and are expected to be the most favorable for numerical computations. Finally, RBS3 and DBS3 probe the strong-field regime, characterized by a highly localized scalar field, generally expected to be the most challenging. The configuration RBS3 is selected on the third branch of solutions, which all three solvers are able to reach. However, as discussed above, the higher branches of DBSs cannot currently be accessed by all three solvers and therefore, DBS3 is chosen near the end of the fundamental branch, just before the backbending.

%
%

\section{Tests and Comparison}

\label{sec:results}




\subsection{Error estimates}\label{sec:err_estim}

Internal convergence criteria differ substantially between the three numerical frameworks and therefore cannot serve as a common measure of accuracy. Instead, we compare the codes through physical, code-independent diagnostics, namely, identities that any exact solution of the EKG system must satisfy, but which a discretized solution fulfills only up to a residual controlled by the numerical resolution. Specifically, for all three solvers, we verify the equivalence between the ADM and Komar definitions of the mass and angular momentum, as well as a virial identity as defined below.

Crucially, none of these relations is imposed during the construction of the solutions; hence, each of them provides an independent measure of the numerical accuracy. Additional identities, such as the constraint equations \eqref{eq:cons} or the absence of conical singularities \eqref{eq:conic_sing}, are also monitored within certain codes, but are not used as common error estimates for solver-specific reasons that make them less relevant across all three approaches. In addition, we have performed standard convergence tests with respect to the resolution.

The first error estimates we consider are
the relative deviations between the Arnowitt-Deser-Misner (ADM) and
Komar definitions of the mass and angular momentum,
\begin{equation}\label{eq:errorsMass}
    \left|\frac{\Delta M}{M}\right|\equiv\left|1-\frac{M_{\rm ADM}}{M_{\rm Komar}}\right|,\quad\quad \left|\frac{\Delta J}{J}\right|\equiv\left|1-\frac{J_{\rm ADM}}{J_{\rm Komar}}\right|,
\end{equation}
where $M_{\rm ADM}$ and $J_{\rm ADM}$ are read off from the the metric functions at spatial infinity, Eq.~\eqref{asym}, while $M_{\rm Komar}$ and
$J_{\rm Komar}$ are computed as volume integrals, Eqs.~\eqref{Mpsi}
and \eqref{JQrel}-\eqref{Q-int}. These two routes probe different regions of the
spacetime: the ADM quantities can be viewed as boundary integrals which are sensitive only to the
asymptotic decay, whereas the Komar quantities accumulate contributions from the entire
matter distribution. For a stationary, asymptotically flat solution of the field equations,
the two necessarily coincide; by virtue of Stokes' theorem, this equivalence is in fact a
consequence of the Einstein equations being satisfied throughout the bulk. Comparing
$M_{\rm ADM}$ with $M_{\rm Komar}$, and likewise for the angular momentum, therefore
provides a test for the numerical quality of the solutions which is global in nature.

As a second and independent diagnostic, we evaluate a relativistic virial identity
derived in Ref.~\cite{Gervalle2025},
\begin{eqnarray}
\label{virial}
    v_\phi=\frac{1}{4\pi}\int_\Sigma T^{\mu\nu}\nabla_{\mu}\zeta_\nu\,\sqrt{-g}\,d^3 x,
\end{eqnarray}
where $\zeta_\mu dx^\mu=r\,e^{F_1}dr$ is a 1-form generating spatial dilatation. This identity is the curved-spacetime generalization of Deser's principal pressure-balance condition for flat-space solitons~\cite{Deser1976}, itself equivalent to the classical scaling relation of Derrick~\cite{Derrick1964}. Since virial identities in relativistic gravity can also be viewed as integrals of suitable combinations of the field equations~\cite{Herdeiro2022}, monitoring the numerical value of $|v_{\phi}|$ provides an additional diagnostic of the overall solution accuracy.

For the ansatz employed in this work, the explicit form of the virial identity
(\ref{virial}) reads,
\begin{eqnarray}
    \nonumber
   v_\phi=&\int_0^\infty dr\int_0^{\pi/2} d\theta\,r^3\sin\theta\left\{e^{-F_0+F_1+F_2}\left[\left(\omega-m W\right)^2\left(\frac{3}{r}-F_{0,r}+F_{1,r}+F_{2,r}\right)-2m\left(\omega-m W\right)W_{,r}\right]\phi^2\right.\\
    \nonumber
    &-e^{F_0+F_1-F_2}\frac{m^2}{r^2\sin^2\theta}\left(\frac{1}{r}+F_{0,r}+F_{1,r}-F_{2,r}\right)\phi^2-e^{F_0+F_1+F_2}\left(\frac{3}{r}+F_{0,r}+F_{1,r}+F_{2,r}\right)\mu^2\phi^2\\
    &\left.-e^{F_0-F_1+F_2}\left[\left(\phi_{,r}^2+\frac{\phi_{,\theta}^2}{r^2}\right)\left(\frac{1}{r}+F_{0,r}+F_{2,r}\right)+\phi_{,r}^2F_{1,r}-\frac{1}{r^2}\phi_{,\theta}^2F_{1,r}+\frac{2}{r^2}\phi_{,r}\phi_{,\theta}F_{1,\theta}\right]\right\}=0~.
\end{eqnarray}

The results of these tests for the six reference configurations are collected in
Tables~\ref{table_spin_newt}--\ref{table_dip_075}. Overall, the three solvers, based
respectively on finite differences (\texttt{FIDISOL/CADSOL}), finite elements
(\texttt{FreeFem}) and spectral methods (\texttt{Kadath}), agree on the global observables
$M$, $J$ (or $Q$) and the central lapse $-g_{tt}(r=0)$ to typically eight 
significant
digits, while the specific error indicators $|\Delta M/M|$, $|\Delta J/J|$ and $|v_\phi|$
range from $10^{-8}$ down to $10^{-15}$. 
However, we emphasize that the resolutions used to obtain these results are substantially higher than those usually adopted in the literature. Here, we deliberately pushed each solver close to the limits of its achievable accuracy; the resulting data may therefore provide useful benchmark values for future numerical studies. In the following, we discuss the results in each of the three solution regimes separately.

For the weakly-relativistic configurations RBS1 and DBS1, the numerical difficulty is the large spatial extent of the scalar field, which must be resolved out to many times its Compton wavelenghth (set to unity in our numerical setups).
Moreover, since we use a radial compactification, the solutions profiles exhibit steep gradients near the asymptotic boundary, $x=1$. This is handled
by increasing the compactification parameter $c$ in Eq.~\eqref{comp} above unit value, which redistributes the resolution towards the far-field region. This strategy yields excellent accuracies for both RBS1 and DBS1, with $|\Delta M/M|,\,|\Delta J/J|,\,|v_\phi|\simeq 10^{-14}-10^{-9}$. The agreement between the three solvers in the global observables is slightly lower for DBS1 (seven to eight significant digits), than for RBS1 (eight to nine significant digits).

The configurations RBS2 and DBS2 ($\omega/\mu=0.8$ and $0.9$, respectively) lie in the maximum-mass region of the fundamental branch, where the solutions are expected to be 
numerically well behaved. 
Accordingly, one anticipates the error indicators to attain their smallest values for these configurations. However, comparing Table~\ref{table_spin_newt} with Table~\ref{table_spin_norm} and Table~\ref{table_dip_099} with Table~\ref{table_dip_075}, we observe that all solvers achieve similarly high accuracies in both the weak- and intermediate-gravity regimes. This performance owes to the optimization of the compactification parameter adopted in our codes, which proves very effective to improve accuracy in numerically-demanding regimes. For RBS2, we obtain $|\Delta M/M|,\,|\Delta J/J|,\,|v_\phi|\simeq 10^{-14}-10^{-9}$, while for DBS2, $|\Delta M/M|,\,|v_\phi|\simeq 10^{-15}-10^{-9}$, comparable to the values obtained for RBS1 and DBS1. At the same time, the three solvers agree to at least eight significant digits for all computed global quantities of both RBS2 and DBS2.


The last two configurations are in the strong-gravity regime: the most demanding one. The configuration RBS3 lies on the third branch of the mass-frequency spiral (left panel of Fig.~\ref{fig_RBS}), deep in
the highly-relativistic region. Here, the central lapse approaches zero (see Fig.~\ref{fig_lapse}, left panel) and the scalar field profile is strongly localized near the origin. 
Accordingly, we adopt the reverse strategy as in the weakly-relativistic case of RBS1: we decrease the compactification parameter $c$ below unit value, which redistribute the resolution towards the origin. Again, this strategy allows us to keep the numerical errors under control, yielding excellent accuracies:
$|\Delta M/M|,\,|\Delta J/J|\simeq10^{-10}-10^{-8}$ and $|v_\phi|\simeq 10^{-13}-10^{-10}$. The smallest errors are obtain with the \texttt{FreeFem}-based code, suggesting that the finite element approach is particularly well-suited for resolving steep field gradients. Despite the increased numerical difficulty, the three solvers still agree to six significant digits, demonstrating a high level of consistency even in extreme regime.

As explained in Sec.~\ref{sec:gen_properties}, the higher branches of dipolar configurations are the most challenging to obtain numerically. To avoid the delicate part of the parameter space where not all solvers are able to obtain solutions, we choose the strong-gravity configuration DBS3 near the end of the fundamental branch (see Fig.~\ref{fig_DBS}). As shown in the right panel of Fig.~\ref{fig_lapse}, the minimal lapse value is $\simeq0.46$, remaining well above zero. Note that this minimum deviates from the central value, $\sqrt{-g_{tt}(r=0)}\simeq 0.61$, as the localization of the field configuration is at a finite distance from the origin along the symmetry axis, at $z_0\simeq 1.7$, rather than near the origin as for RBS3. This peculiar spatial localization makes the tuning of the compactification parameter inefficient in this case. Nevertheless, our numerical results remain highly accurate, with $|\Delta M/M|\simeq 10^{-15}-10^{-9}$ and $|v_\phi|\simeq 10^{-12}-10^{-11}$. The three solver furthermore agree to eight significant digits for the mass and Noether charge, and to nine digits for the lapse.

A feature specific to the dipolar stars is the coordinate separation $z_{\rm d}$ of the two
scalar lumps, defined as twice the location $z_0$ of the maximum of $\phi$ along the symmetry axis.
As $\omega$ decreases from $\mu$ -- where the lumps are located arbitrarily far from the origin, $z_{\rm d}\to\infty$ -- towards stronger gravity, the lumps migrate
inwards, and $z_{\rm d}$ shrinks from $\simeq 47$ (DBS1) to $\simeq 11$ (DBS2) and
$\simeq 3.4$ (DBS3) (see also Fig.~5 in Ref.~\cite{Cunha:2022tvk}). We note that $z_{\rm d}$ is reproduced across the solvers only to about five
significant
digits, which is noticeably less than the eight to nine digits achieved for
$M$, $Q$ and the central lapse. This is to be expected, since $z_{\rm d}$ is not a global
integral but a 
coordinate-dependent quantity, obtained by locating a maximum through interpolating the
discrete profile; it is therefore sensitive not only to the resolution, but also to the interpolation order and maximization method employed in the post-processing.

\begin{figure}[ht]
\includegraphics[width=0.48\linewidth]{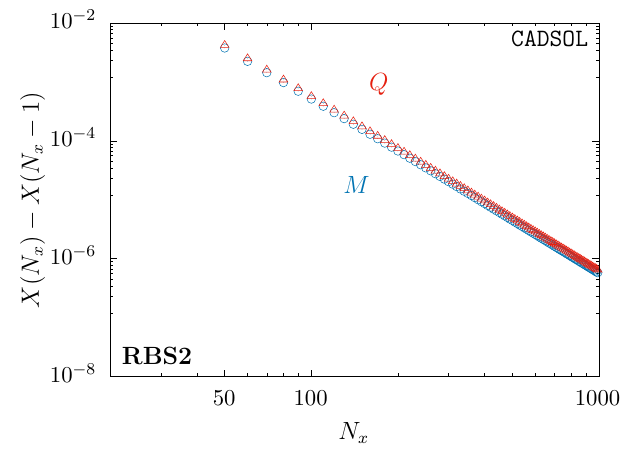}
\qquad{}
\includegraphics[width=0.48\linewidth]{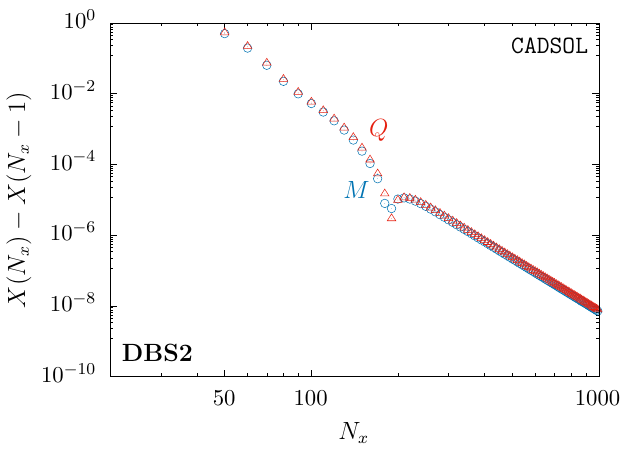}
\caption{\small 
{\it Convergency test -- \texttt{CADSOL}.}
The convergency of the numerically obtained values for $M$ and $Q$ when increasing $N_x$ is shown with fixed $N_\theta=100$, for the RBS2 solution (left) and the DBS2 solution (right). The data points approach a slope of $\sim -3$.}
\label{conv_cadsol}
\end{figure}

\begin{figure}[ht]
\includegraphics[width=0.48\linewidth]{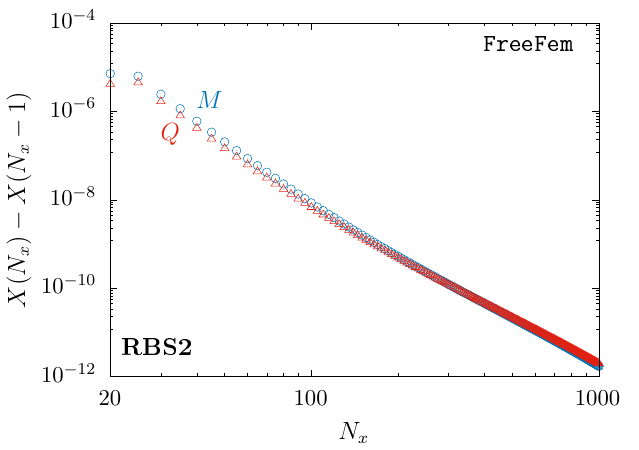}
\qquad{}
\includegraphics[width=0.48\linewidth]{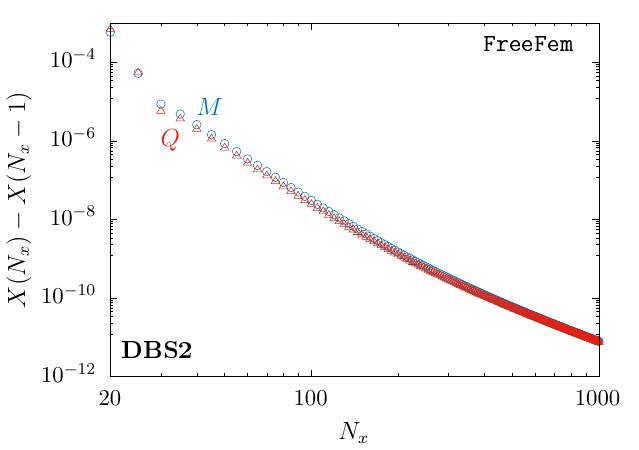}
\caption{\small 
{\it Convergency test -- FreeFem.}
The convergency of the numerically obtained values for $M$ and $Q$ when increasing $N_x$ is shown with fixed $N_\theta=50$, for the RBS2 solution (left) and the DBS2 solution (right). The data points approach a slope of $\sim -4$.}
\label{conv_freefem}
\end{figure}

\begin{figure}[ht]
\centering
\includegraphics[width=0.47\linewidth]{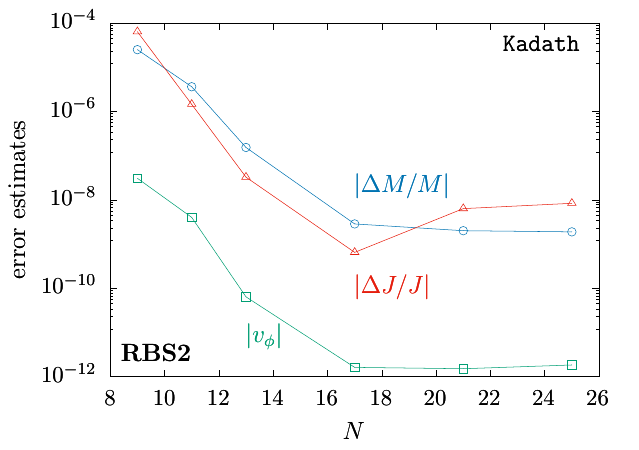}
\qquad{}
\includegraphics[width=0.47\linewidth]{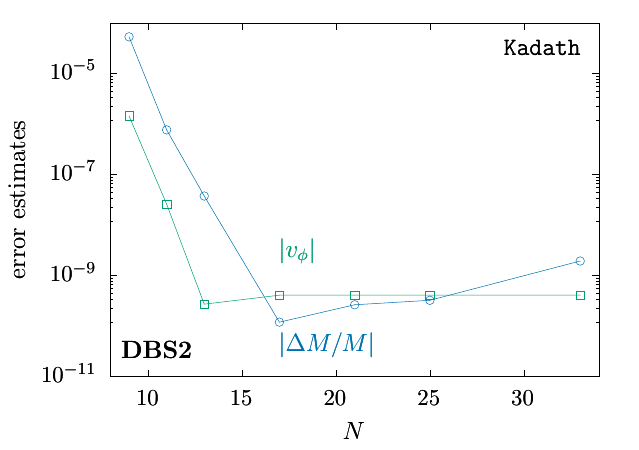}
\caption{\small 
{\it Convergence test -- Kadath.} 
The convergence of error estimates obtained for the RBS2 solution (left) and the DBS2 solution (right). $N$ is the number of spectral coefficients in both $\theta$ and $r$ (in each spherical domain).}
\label{conv_kadath}
\end{figure}

Finally, we present in Figs.~\ref{conv_cadsol}-\ref{conv_kadath} the convergence tests for the three solvers obtained by systematically increasing the resolution for the reference configurations RBS2 and DBS2. We observe for both \texttt{CADSOL} (Fig.~\ref{conv_cadsol}) and \texttt{FreeFem} (Fig.~\ref{conv_freefem}) the power-law convergence characteristic of finite difference and finite element methods, as seen on the log-log plots, with slopes of $\sim -3$ and $\sim -4$, respectively. By contrast, \texttt{KADATH} exhibits the exponential convergence typical to spectral methods, as evidenced by the linear behavior on the semi-log plots shown in Fig.~\ref{conv_kadath} where only the vertical axes are displayed on a logarithmic scale.


\begin{table}
    \centering
    \renewcommand{\arraystretch}{1.3}
    \begin{tabular}{ |c||c|c|c|c|c|c|  }
        \hline
        \textbf{RBS1} & $M_{\rm ADM}$  & $J_{\rm ADM}$ & $-g_{tt}(r=0)$ & $|\Delta M/M|$ & $|\Delta J/J|$ & $|v_\phi|$ \\
        \hline
        CADSOL & $0.29358860521$ & $ 0.29407622678$ & $ 0.98368064137$ & $2.76\times10^{-10}$ & $9.98\times10^{-9\hphantom{0}}$ &  $3.81\times 10^{-10}$\\
         \hline
        FreeFem & $0.29358860531$ & $0.29407622392$ & $0.98368064144$ & $6.72\times 10^{-14}$ & $9.21\times 10^{-11}$ & $4.79\times 10^{-14}$ \\
        \hline
        Kadath  & $0.29358860501$ & $0.29407622349$ & $0.98368064147$ & $7.54\times 10^{-9\hphantom{0}}$  & $5.07\times10^{-9\hphantom{0}}$   & $5.06\times 10^{-14}$  \\
        \hline
        \end{tabular}
    \caption{Output parameters for the single rotating boson star with azimuthal number $m=1$ and frequency $\omega=0.995$, belonging to the fundamental branch (Newtonian regime).}
    \label{table_spin_newt}
\end{table}

\begin{table}
    \centering
    \renewcommand{\arraystretch}{1.3}
    \begin{tabular}{ |c||c|c|c|c|c|c|  }
        \hline
        \textbf{RBS2} & $M_{\rm ADM}$  & $J_{\rm ADM}$ & $-g_{tt}(r=0)$ & $|\Delta M/M|$ & $|\Delta J/J|$ & $|v_\phi|$ \\
        \hline
        CADSOL & $1.30778592527$ & $1.37181314851$ & $0.41992815680 $ & $6.54\times10^{-9\hphantom{0}}$  & $5.35\times10^{-10}$  & $1.37\times 10^{-9\hphantom{0}}$ \\
        \hline
        FreeFem & $1.30778591707$ & $1.37181309885$ & $0.41992816144$ & $5.00\times 10^{-11}$ & $6.86\times 10^{-14}$ & $3.67\times 10^{-11}$ \\
        \hline
        Kadath  & $1.30778591498$ & $1.37181309819$ & $0.41992816299$ & $2.84\times10^{-9\hphantom{0}}$ & $6.49\times10^{-10}$ & $1.62\times10^{-12}$\\
        \hline
        \end{tabular}
    \caption{Output parameters for the single rotating boson star with azimuthal number $m=1$ and frequency $\omega=0.8$, belonging to the fundamental branch (``normal'' regime).}
    \label{table_spin_norm}
\end{table}

\begin{table}
    \centering
    \renewcommand{\arraystretch}{1.3}
    \begin{tabular}{ |c||c|c|c|c|c|c|  }
        \hline
        \textbf{RBS3} & $M_{\rm ADM}$  & $J_{\rm ADM}$ & $-g_{tt}(r=0)$ & $|\Delta M/M|$ & $|\Delta J/J|$ & $|v_\phi|$ \\
        \hline
      CADSOL & $0.69825745423$ & $0.46023166741$ & $ 1.5803761\times 10^{-4}$ & 
      $9.21 \times 10^{-8\hphantom{0}}$ & 
      $4.56 \times 10^{-8\hphantom{0}}$ & $2.20\times 10^{-10}$ \\
        \hline
        FreeFem & $0.69825731812$ & $0.46023151056$ & $1.5803920\times 10^{-4}$ & $2.60\times 10^{-10}$ & $2.43\times 10^{-10}$ & $4.15\times 10^{-13}$ \\
        \hline
        Kadath  & $0.69825735860$ & $0.46023155546$ & $1.5803969\times 10^{-4}$ & $4.57\times10^{-9\hphantom{0}}$ & $1.02\times10^{-8\hphantom{0}}$ & $2.31\times10^{-12}$\\
        \hline
        \end{tabular}
    \caption{Output parameters for the single rotating boson star with azimuthal number $m=1$ and frequency $\omega=0.9$, belonging to the third branch (strong gravity regime).}
    \label{table_spin_strong}
\end{table}
\begin{table}
    \centering
    \renewcommand{\arraystretch}{1.3}
    \begin{tabular}{ |c||c|c|c|c|c|c|  }
        \hline
        \textbf{DBS1} & $M_{\rm ADM}$  & $Q$ & $-g_{tt}(r=0)$ & $z_{\rm d}$ & $|\Delta M/M|$ & $|v_\phi|$ \\
        \hline
        CADSOL &  $0.37309319323$ &  $0.37432320764$ & $0.97244713482$ & $47.1113$ & $3.48\times 10^{-9\hphantom{0}}$ & $2.30\times 10^{-11}$ \\
        \hline
        FreeFem & $0.37309320428$ &  $0.37432321926$ & $0.97244713269$ & $47.1121$ & $4.28\times 10^{-13}$ & $7.38\times 10^{-14}$ \\
        \hline
        Kadath & $0.37309319695$ & $0.37432321138$ & $0.97244713457$ & $47.1120$ & $1.37\times10^{-9\hphantom{0}}$ & $2.22\times10^{-12}$\\
        \hline
        \end{tabular}
    \caption{Output parameters for the dipolar boson star with azimuthal number $m=0$ and frequency $\omega=0.99$, belonging to the fundamental branch.}
    \label{table_dip_099}
\end{table}

\begin{table}
    \centering
    \renewcommand{\arraystretch}{1.3}
    \begin{tabular}{ |c||c|c|c|c|c|c|  }
        \hline
        \textbf{DBS2} & $M_{\rm ADM}$  & $Q$ & $-g_{tt}(r=0)$ & $z_{\rm d}$ & $|\Delta M/M|$ & $|v_\phi|$ \\
        \hline
        CADSOL &  $0.96135059489$&$0.98843923079$ &$0.73790843989$ & $11.3321$ & $2.55\times 10^{-9\hphantom{0}}$ & $3.11 \times 10^{-12}$ \\
        \hline
        FreeFem & $0.96135059488$&$0.98843923129$ & $0.73790843937$ & $11.3323$ & $1.67\times 10^{-15}$ & $1.91\times 10^{-11}$ \\
        \hline
        Kadath &  $0.96135059385$&$0.98843923038$ & $0.73790844000$ & $11.3321$ & $1.20\times10^{-10}$ & $4.09\times10^{-10}$ \\
        \hline
        \end{tabular}
    \caption{Output parameters for the dipolar boson star with azimuthal number $m=0$ and frequency $\omega=0.9$, belonging to the fundamental branch.}
    \label{table_dip_09}
\end{table}
\begin{table}
    \centering
    \renewcommand{\arraystretch}{1.3}
    \begin{tabular}{ |c||c|c|c|c|c|c|  }
        \hline
        \textbf{DBS3} & $M_{\rm ADM}$  & $Q$ & $-g_{tt}(r=0)$ & $z_{\rm d}$ & $|\Delta M/M|$ & $|v_\phi|$ \\
        \hline
        CADSOL  &$0.90086400878$&$0.89927212269$ & $0.37183577451$ & $3.39915$& $5.45\times 10^{-9\hphantom{0}}$& $8.03\times 10^{-11}$\\
        \hline
        FreeFem &$0.90086400713$&$0.89927212689$ & $0.37183577483$ & $3.39919$ & $2.22\times 10^{-15}$ & $3.62\times 10^{-11}$ \\
        \hline
        Kadath & $0.90086400667$&$0.89927212668$ & $0.37183577499$ & $3.39915$ & $3.32\times10^{-10}$ & $9.77\times10^{-12}$ \\
        \hline
        \end{tabular}
    \caption{Output parameters for the dipolar boson star with azimuthal number $m=0$ and frequency $\omega=0.75$, belonging to the fundamental branch.}
    \label{table_dip_075}
\end{table}

\subsection{Running time}\label{sec:runtime}

Besides the numerical accuracy assessed through the code-independent
diagnostics introduced in the previous subsection, computational cost provides
an important practical criterion for comparing the three solvers. The runtime
comparisons reported in Tables~\ref{table_runtime} and
\ref{table_runtime2} were carried out by running all three codes on the same
desktop machine, equipped with a 12th-generation Intel Core i5-12400 processor
(6 physical cores, 12 threads, and a clock frequency up to $4.4$~GHz)
. The runtime benchmarks used the same six physical CPU cores for each solver. \texttt{FIDISOL/CADSOL} ran as a single process with six OpenMP threads, whereas \texttt{FreeFem} and \texttt{Kadath} each used six MPI processes, 
with each MPI process restricted to one computational thread.
As a representative case, we consider the RBS2 reference solution ($m=1$,
$\omega=0.8$) and measure, for each code, the wall-clock time required to obtain
it starting from a converged initial guess at the nearby frequency
$\omega=0.799$ (i.e. a step of $\Delta\omega=0.001$).

Beyond the use of identical hardware, a meaningful runtime comparison requires
specifying which quantity is held fixed across the solvers. Since the number of degrees
of freedom, that is, the number of algebraic unknowns effectively solved by each code, has a different meaning across the three discretization schemes, 
physical accuracy is the natural quantity to hold fixed, and we consider two complementary benchmarks. In the first one (Table~\ref{table_runtime2}),
the resolution of each solver is adjusted so as to achieve the same prescribed
physical accuracy, allowing the codes to be compared on an equal practical footing. In
the second one (Table~\ref{table_runtime}), we report the runtimes associated
with resolutions used for the high-accuracy solutions of Table~\ref{table_spin_norm}; in this case, neither the accuracy nor the number of degrees of freedom is fixed, but the resulting timings quantify the cost of obtaining the high-accuracy results reported in this work on the same hardware.

In Table~\ref{table_runtime2}, we adjusted the resolution of each solver so
that the global error indicators on the mass and angular momentum, as measured from Eq.~\eqref{eq:errorsMass}, is of order $10^{-5}$, which is representative of the typical accuracy
targeted in the literature. The \texttt{FreeFem} has the shortest wall-clock time in this benchmark, with a runtime of $14\,{\rm s}$, compared with $2\,{\rm min}$ for
the two other solvers. Interestingly, keeping the error at this level requires
only $N_{\rm DoF}=40\,905$ degrees of freedom for the finite element code,
whereas the finite difference code requires $N_{\rm DoF}=100\,000$, more than
twice as many. The spectral code requires the fewest degrees of freedom,
$N_{\rm DoF}=5\,265$, consistently with its exponential convergence.

\begin{table}[ht]
    \centering
    \renewcommand{\arraystretch}{1.3}
    \begin{tabular}{ |c||c|c|c| }
        \hline
        \textbf{Solver} & Resolution & $N_{\rm DoF}$ & Wall-clock time \\
        \hline
        CADSOL  & $N_x\times N_\theta = 200\times100$ & $100\,000$ & $2 \text{ min}$ \\
        \hline
        FreeFem & $N_x\times N_\theta = 40\times 50$ & $40\,905$ & $14 \text{ s}$ \\
        \hline
        Kadath  & $n_{\rm dom}\times N_x\times N_\theta = 13\times 9\times 9$ & $5\,265$ & $2 \text{ min}$ \\
        \hline
    \end{tabular}
    \caption{Wall-clock time required to compute the RBS2 reference solution
    ($m=1$, $\omega=0.8$) with each solver, starting from a converged initial
    guess at $\omega=0.799$. For each code, the resolution is adjusted such
    that the error indicators on the mass and angular momentum are of order $10^{-5}$,
    $|\Delta M/M|\simeq|\Delta J/J|\simeq 10^{-5}$, which is representative of the typical
    accuracy targeted in the literature. The corresponding total number of
    degrees of freedom, $N_{\rm DoF}$, is also given.}
    \label{table_runtime2}
\end{table}

For the runtimes of Table~\ref{table_runtime}, we use for each solver the
resolution that produced the corresponding solution reported in
Table~\ref{table_spin_norm}; the comparison is therefore performed at the maximum accuracy reported here rather than at a fixed accuracy or a fixed number of degrees of freedom. 
Remarkably, the runtime for \texttt{FreeFem} is
again substantially shorter than for the two other solvers, by a factor of two compared with \texttt{Kadath} and by a factor of six compared with
\texttt{CADSOL}. This is even more striking given that the total number of
degrees of freedom effectively used for this run is larger for
\texttt{FreeFem} ($N_{\rm DoF}=1\,375\,035$) than for \texttt{CADSOL}
$N_{\rm DoF}=500\,000$) and \texttt{Kadath} ($N_{\rm DoF}=18\,785$). 

Finally, comparing the two benchmarks shows that reducing the error indicators from the $10^{-5}$ level down to the
$10^{-14}-10^{-9}$ level of Table~\ref{table_spin_norm}, 
the runtime increases by a factor of approximately $5$ for \texttt{Kadath}, $14$ for \texttt{CADSOL}, and $20$ for \texttt{FreeFem}. At the same time, in each case, the number of degrees of freedom has increased by a factor $\sim 3$, $\sim 5$ and $\sim 30$, respectively. Despite exhibiting by far the largest increase in degrees of freedom, the \texttt{FreeFem}-based code thus achieve the most favorable runtime scaling, followed by \texttt{Kadath} and \texttt{CADSOL}.

\begin{table}[ht]
    \centering
    \renewcommand{\arraystretch}{1.3}
    \begin{tabular}{ |c||c|c|c| }
        \hline
        \textbf{Solver} & Resolution & $N_{\rm DoF}$ & Wall-clock time \\
        \hline
        CADSOL  & $N_x\times N_\theta = 1000\times100$ & $500\,000$ & $28 \text{ min}$ \\
        \hline
        FreeFem & $N_x\times N_\theta = 420\times 163$ & $1\,375\,035$ & $5 \text{ min}$ \\
        \hline
        Kadath  & $n_{\rm dom}\times N_x\times N_\theta = 13\times17\times 17$ & $18\,785$ & $10 \text{ min}$ \\
        \hline
    \end{tabular}
    \caption{Same as Table~\ref{table_runtime2}, but with the resolution of
    each code being that of the corresponding entry in
    Table~\ref{table_spin_norm}: the comparison is thus performed at the
    maximum accuracy reported in this work rather than at a fixed accuracy or
    a fixed number of degrees of freedom.}
    \label{table_runtime}
\end{table}

 \subsection{Specific aspects}\label{sec:specs}

 \subsubsection{CADSOL}
 \label{spec_cadsol}

As previously discussed, \texttt{CADSOL} is written in \texttt{Fortran 90} and follows
the finite-difference strategy described in Sec.~\ref{sec:fidisol_cadsol}. The code is
organized into separate routines for the construction of the computational grid, the
initialization of the fields, the evaluation of the PDE residuals, the implementation of
the boundary conditions, the assembly of the analytical Jacobian, the Newton-Raphson
iteration, while the post-processing of physical quantities is performed with
\textit{Mathematica} using the solver outputs. The explicit form of the field
equations and of the corresponding Jacobian matrices were derived symbolically with
\textit{Mathematica} and then translated into \texttt{Fortran 90} routines.

Owing to the symmetry properties of the configurations with respect to the equatorial plane, the computational domain is restricted to $[0,\infty)\times[0,\pi/2]$. 
To avoid the use of a cutoff radius, we 
use the compactified radial coordinate $x$ as defined in Eq. (\ref{comp}),
such that  the semi-infinite interval $r\in[0,\infty)$ is mapped to the finite segment $x\in[0,1]$. We then make the necessary substitutions in the field equations,
	\begin{equation}
    \label{eq:substitute}
		\mathcal{F}_{, r} \longrightarrow \frac{1}{c}(1-x)^2 \mathcal{F}_{, x}, \quad \mathcal{F}_{, r r} \longrightarrow \frac{1}{c^2}(1-x)^4 \mathcal{F}_{, x x}-\frac{2}{c^2}(1-x)^3 \mathcal{F}_{, x} \ .
	\end{equation}
    
In the computations reported here, the \texttt{CADSOL} grid is taken to be equidistant in the
numerical coordinates~\((x,\theta)\). 
The physical radial coordinate $r$ is therefore
nonuniformly sampled even when the numerical grid is uniform in \(x\). The compactification parameter
\(c\) in Eq. (\ref{comp}) controls how the grid resolution is distributed in physical space and is typically set to $1$. Larger values of
\(c\) move the far-field region toward smaller values of \(x\), which is useful for weakly
relativistic configurations with extended scalar profiles. Smaller values of \(c\) increase
the effective resolution near the origin and can improve the description of strongly
localized configurations. Thus, \(c\) is treated as an input numerical parameter and adjusted
for each branch, or for each benchmark solution, in order to optimize the accuracy of
the computed global quantities and minimize the error indicators. Although \texttt{CADSOL} also allows nonequidistant grids in the numerical coordinates, the uniform \((x,\theta)\) grids used
here provide a transparent setting for convergence tests.

The nonlinear system is solved using the Newton-Raphson method, starting from an initial guess.
A converged
solution at a given scalar-field frequency \(\omega\) is used as an initial guess for a
nearby value of \(\omega\), typically with frequency steps of order
\(\Delta\omega\simeq 10^{-3}\) for the benchmark sequences. At each Newton step, the
linear system for the correction is solved using LINSOL.
%
In the present calculations, the method using a smoothed bi-conjugate gradient, \textsc{BICO}, has been the most robust choice, though other methods in the bundle may be preferable for specific problems.  Before iteration, LINSOL can scale the linear system to improve conditioning and component balance. By default we use \emph{row normalization} (sum of absolute values per row equal to one).
The Newton iteration is stopped once the algebraic residual is compatible with the
chosen tolerance and with the discretization-error estimate obtained from the
difference-of-formulae procedure described in Sec.~\ref{sec:fidisol_cadsol}.

The version of \texttt{FIDISOL/CADSOL} used in our tests also includes a shared-memory
parallelization based on OpenMP. This is not a domain-decomposition
parallelization: the numerical domain is not divided into independent subdomains,
and no MPI communication is used. Instead, the original finite-difference data
structure is kept, and OpenMP is used to parallelize the most expensive loops
over grid points, difference stars, matrix diagonals, and vector entries. In
practice, the parallel part is mainly associated with the construction of the
finite-difference formulae, the computation of derivative and error stencils, the
assembly of the discretized Jacobian matrix, and the vector operations appearing
in the iterative solution of the Newton correction. The Newton continuation
itself, the input/output operations, and the global convergence tests remain
serial, or require synchronization. Thus, the OpenMP version should be understood
as a node-level acceleration of the original code, rather than as a fully
distributed parallel solver.

This distinction is important for the expected computational behaviour. Since the
domain is not split among processors, increasing the number of threads does not
reduce the memory footprint of the problem, nor does it change the size of the
linear system solved at each Newton step. The gain comes only from sharing the
work of the loop-based kernels. Therefore, one should not expect an ideal linear
speedup with the number of threads. For small grids, the overhead of creating and
synchronizing threads can be comparable to the cost of the parallel loops, and
the improvement may be modest. For larger grids, the speedup should become more
visible, because the stencil construction, Jacobian assembly, and matrix-vector
operations contain enough independent work to keep several cores active. However,
as the number of threads is increased further, the scaling is expected to
saturate.

The code has modest software requirements. It can be compiled with a standard
\texttt{Fortran 90} compiler and we have
tested the code with both Intel Fortran compilers, \texttt{ifort} and
\texttt{ifx}. The raw output consists of simple text files containing the grid
values of the fields and diagnostic error indicators. These files are post-processed with \textit{Mathematica} to compute derived
quantities and locate the maxima of the dipolar scalar-field profiles.

The number of algebraic unknowns in a \texttt{CADSOL} run is,
\begin{equation}
    N_{\rm DoF}=N_K\,N_x\,N_\theta ,
    \label{eq:dof_cadsol}
\end{equation}
where \(N_K=5\) for rotating boson stars and \(N_K=4\) for static dipolar boson stars.
This degree-of-freedom count is the most meaningful quantity for runtime comparisons with
finite-element and spectral implementations, as the interpretation of the resolution
parameters differs among the three numerical methods.

The convergence of the \texttt{CADSOL} implementation was tested by increasing the radial
resolution \(N_x\) while keeping the angular resolution fixed. Figure~\ref{conv_cadsol}
shows the convergence of the mass \(M\) and charge \(Q\) for the RBS2 and DBS2
reference solutions at fixed \(N_\theta=100\). The differences,
\begin{equation}
    X(N_x)-X(N_x-1), \qquad X\in\{M,Q\},
    \label{eq:conv_estimator}
\end{equation}
approach an approximate slope of \(-3\) on a log--log scale, consistently with the
finite-difference order used in these runs. 

For orientation, a representative RBS2 computation converges in four Newton
iterations and takes 
\textit{typically} two minutes on a standard PC, although the precise runtime significantly depends on the grid size, the compiler, the OpenMP configuration, and the
chosen tolerance. 

The physical quantities are computed \textit{a posteriori} in \textit{Mathematica} from the
\texttt{CADSOL} output. The Komar mass, Noether charge, angular momentum and virial
identity are evaluated through the corresponding volume integrals, after
interpolating the numerical functions and their derivatives. The ADM mass
and angular momentum are instead extracted from the asymptotic behaviour of
the metric functions. Comparing the metric ansatz \eqref{metric_ansatz} with the
asymptotically flat expansion \eqref{AF-expansion}, and using the compactified coordinate
$x$ in \eqref{comp}, one obtains
(we recall the numerics is done with $W=\Omega/r$)
\begin{equation}
 M_{\rm ADM}
 =
 \frac{c}{N_\theta}
 \sum_{j=1}^{N_\theta}
 \left.
 \partial_x F_0(x,\theta_j)
 \right|_{x=1},
 \qquad
 J_{\rm ADM}
 =-
 \frac{c^2}{4 N_\theta}
 \sum_{j=1}^{N_\theta}
 \left.
 \partial_x^2 \Omega(x,\theta_j)
 \right|_{x=1}.
 \label{eq:CADSOL-ADM}
\end{equation}
The averages are taken over the angular grid points at infinity ($x=1$), the values of the derivative being  directly provided by the solver. For an exact
asymptotically flat solution, the quantities entering these sums are
independent of $\theta$; the averaging therefore reduces the residual
finite-grid fluctuations.

The \texttt{CADSOL} package is not distributed as a public
package, unlike the FreeFem and Kadath codes used in the comparison. A related version of the \texttt{CADSOL} infrastructure can also be used for genuinely three-dimensional elliptic problems, with the same basic organization: residual evaluation, analytical Jacobian construction, Newton iteration, linear solution with LINSOL, and post-processing of the physical observables.

\subsubsection{FreeFem}
\label{spec_freefem}


To run our finite element solver, \texttt{FreeFem} must be installed on the computer. The sources are publicly available on GitHub\footnote{\url{https://github.com/FreeFem/FreeFem-sources}; documentation and examples can be found at \url{https://freefem.org/}} and can be compiled with a standard C++ compiler by following the instructions in the official documentation. Parallel execution is supported by \texttt{MPICH}. As with \texttt{FIDISOL/CADSOL}, the parallelization strategy is not a \textit{true} domain decomposition\footnote{Full domain-decomposition parallelization is available in \texttt{FreeFem} via the \texttt{PETSc} module, which interfaces with the \texttt{HPDDM} library. However, for the purpose of the present work, which is restricted to effectively 2D problems, the shared-memory parallelization provided by \texttt{MUMPS} is sufficient.}: all processors hold copies of the full set of degrees of freedom, with the mesh elements distributed among processors solely for the purpose of parallelizing the assembly of the Jacobian matrix and the associated residual vector (second term in the right-hand side of Eq.~\eqref{eq:newt_step}). The inversion of the Jacobian is handled independently by the built-in \texttt{MUMPS} parallel sparse solver. This parallelization strategy shares the same limitations as those discussed in Sec.~\ref{spec_cadsol}.      

Our code is written in the \texttt{FreeFem} language and proceeds through the following steps: (i) mesh construction, (ii) definition of the finite element space, (iii) specification of an initial guess, (iv) Newton-Raphson iterations, (v) computation of output quantities, (vi) storage of the converged solution and mesh, and (vii) export of solution profiles and output data to external files. The converged solutions and associated meshes from step (vi) are stored in text files whose precise syntax is handled automatically by \texttt{FreeFem}, allowing them to be read easily by the code in subsequent runs. The solution profiles we export in step (vii) are intended for post-processing only; for convenience, the field values are exported on the nodes of a uniform $(x,\theta)$ grid analogous to those used by \texttt{FIDISOL/CADSOL}, rather than on the mesh (mid-)vertices of the \texttt{FreeFem} solver.

The meshes used to solve the field equations are nonuniform, with increased resolution near selected boundaries. For the configurations RBS1, DBS1, DBS2, and DBS3, we use the mesh shown in the center panel of Fig.~\ref{meshes}, which is refined near $x=1$ and $\theta=0$. For RBS2 and RBS3, the mesh shown in the right panel of of Fig.~\ref{meshes} is adopted, featuring an additional refinement near $x=0$. These targeted refinements were found to significantly improve numerical accuracy. The overall strategy is to have a refined resolution where it is most needed: near the asymptotic boundary where small variations in $x$ correspond to large variations in the physical radial coordinate $r$; and in regions where the field profiles exhibit steep gradients, which typically occur near the symmetry axis ($\theta=0$) and in the vicinity of the origin for the more compact configurations RBS2 and RBS3. The dipolar configurations DBS2 and DBS3 are also compact, but the field localization occurs at a finite distance along the symmetry axis rather than near the origin. In principle, numerical accuracy could be improved for these DBSs by refining the mesh near the maximum of the scalar profile; however, since its location depends on the input frequency $\omega$, implementing such a refinement systematically is nontrivial and is left for future developments.

In addition to mesh refinements, the same compactification parameter $c$ as in Eq.~\eqref{comp} is tuned to optimize numerical accuracy, again depending on the spatial localization of the solutions. In the Newtonian regime, $\omega/\mu\to 1$, the scalar field profile becomes increasingly concentrated near $x=1$. In this case, values $c>1$ are advantageous, as increasing $c$ shifts the far-field region toward smaller values of $x$ (for $x=1/2$, one has $r=c$). In practice, we use $c=2.868$ for RBS1 and $c=12.635$ for DBS1. For intermediate solutions along the first branch, the profiles are less sharply localized and values $c\sim 1$ are optimal; we choose $c=1.825$ for RBS2 and $c=1.619$ for DBS2. For rotating solutions in the strong-gravity regime, the fields are strongly localized near the origin. In this case, choosing $c<1$ mitigate the development of steep gradients near $x=0$; RBS3 is constructed with $c=0.761$. Finally, for dipolar solutions in the strong-gravity regime, such as DBS3, the scalar field maximum is at a finite distance from the origin along the symmetry axis. In this case, choosing values $c<1$ do not improve the numerical accuracy, and we therefore set $c=1$.

The finite element space $V_h$ is chosen as the space of continuous, piecewise polynomial functions of degree $k=2$. This choice provides a balanced compromise between memory cost and numerical accuracy, since the dimension of $V_h$ -- and hence the size of the resulting nonlinear system to solve -- increases with $k$ for a given mesh.

Initial guesses can be specified either using analytical expressions, with \texttt{FreeFem} computing internally the corresponding expansion coefficients in the finite element basis, or by using data from a previously converged solution. In practice, we adopt the latter approach for constructing sequences of solutions in which $\omega$ is varied in small increments, typically of order $\Delta\omega\simeq 10^{-4}-10^{-2}$. The expansion coefficients are then read directly from the saved output files.

Analytical expressions for the weak formulation of the field equations and the corresponding Jacobian required for the Newton-Raphson iterations are derived using \textit{Mathematica} and then translated into \texttt{FreeFem}'s language. As explained in Sec.~\ref{fem}, boundary conditions are enforced through the weak formulation by construction in the finite element approach. For the metric functions, rather than working directly with $F_1$, we introduce the auxiliary field $\Delta F\equiv F_2-F_1$ and impose the absence of conical singularities on the symmetry axis \eqref{eq:conic_sing} by requiring $\Delta F\big|_{\theta=0}=0$. The remaining condition, $\partial_\theta F_1\big|_{\theta=0}=0$, is not imposed in practice. Instead, we verify \textit{a posteriori} that it is satisfied by the converged solutions up to a residual consistent with the code-independent error estimators.

To construct the Jacobian, we introduce small variations $\delta{\cal F}=\{\delta\phi,\delta F_0,\delta F_1,\delta F_2,\delta W\}$ and differentiate the weak equations of the form \eqref{weak_form_1} with respect to the fields ${\cal F}$. The resulting expression, involving ${\cal F}$, the test functions $v_{\cal F}$ and the variations $\delta{\cal F}$, is then implemented in the code, with \texttt{FreeFem} performing the finite-element basis decomposition internally to obtain the Jacobian entries $J_{ij}=\partial E_i/\partial{\cal F}_j$. 
At each iteration, the Jacobian inverse $J^{-1}$ is computed using the built-in \texttt{MUMPS} solver. 

Solutions are usually constructed on meshes with typical resolution $N_x\times N_\theta=130\times 30$ or $130\times50$, for which the runtime per solution typically remains below one minute. 
Note that this is slightly more conservative than in the comparison presented in Table~\ref{table_runtime2}. However, to obtain the high-accuracy results reported in Tables~\ref{table_spin_newt}-\ref{table_dip_075}, we used significantly finer meshes,
resulting in correspondingly longer runtimes (see, for example, the Table~\ref{table_runtime} for RBS2 configuration).

The number of degrees of freedom -- that is, the size of the algebraic system effectively solved by the code -- is,
\begin{equation}
    N_\text{DoF}=\big[4N_x N_\theta+2(N_x+N_\theta)+1\big]\times N_K,
    \label{eq:dof_freefem}
\end{equation}
where $N_K$ is the number of unknown functions. This counts reflects the use of second-order finite elements, which assign degrees of freedom both at mesh nodes and at the midpoints of mesh edges. Comparing with Eq.~\eqref{eq:dof_cadsol}, which simply corresponds to the number of grid nodes in the \texttt{FIDISOL/CADSOL} discretization, one sees that \texttt{FreeFem} employs more than four times as many internal degrees of freedom for the same resolution parameters.

In \texttt{FreeFem}, the finite element representation of the field variables as in Eq.~\eqref{decomp_fe} makes the evaluation of integrals straightforward. Accordingly, all output quantities involving bulk integrals such as $M_\text{Komar}$, $J_\text{Komar}$, $Q$ and $v_\phi$ are computed directly within the code. Asymptotic quantities, including the ADM mass $M_\text{ADM}$ and angular momentum $J_\text{ADM}$, are also computed internally, using the boundary values of $\partial_xF_{0}$ and $\partial_x\tilde{W}\equiv\partial_x(r^2W)$ at $x=1$. In practice, these quantities are obtained from the boundary integrals,
\begin{equation}
\label{eq:surf_adm}
    M_\text{ADM}=c\left.\int_0^{\pi/2}\partial_x F_0\sin\theta\,d\theta\right|_{x=1},\quad J_\text{ADM}=-\frac{c}{2}\left.\int_0^{\pi/2}\partial_x\tilde{W}\sin\theta\,d\theta\right|_{x=1},
\end{equation}
where we have used the asymptotic expansions \eqref{asym} and the fact that $\partial_x F_0,\,\partial_x\tilde{W}$ are $\theta$-independent at $x=1$. The coordinate distance $z_{\rm d}$ for dipolar solutions is determined by locating the maximum of the scalar field $\phi$ along the symmetry axis. This is performed in \textit{Mathematica} using a cubic interpolation of the numerical scalar field profile. We have checked the convergence of all output quantities when the mesh resolution is increased using the same convergence estimator as in Eq.~\eqref{eq:conv_estimator}. This is illustrated in Fig.~\ref{conv_freefem}, where, for fixed $N_\theta=50$ and increasing $N_x$ from 20 to 1000, we obtain a convergence slope approaching $-4$ on a log-log scale, indicating a fourth-order convergence.

The current implementation of the code is restricted to effective 2D problems. In principle, extending the framework to fully 3D configurations is relatively straightforward within \texttt{FreeFem}, which natively supports 3D finite-element discretization. Such an extension is left for future work.

\subsubsection{Kadath}
\label{spec_kadath}


Compilation requires standard scientific libraries together with MPI support for parallel execution. In practice, parallelization is essential to obtain reasonable runtimes and avoid memory exhaustion for high-resolution configurations involving several spectral domains and large numbers of spectral coefficients. Our implementation follows the standard workflow: construction of the multidomain spectral space, specification of coordinate mappings and domain boundaries, definition of the field variables, constants, differential equations and constraints, specification of analytical initial guesses, Newton-Raphson iterations and computation of global quantities.

The field equations are implemented directly using the very simple \texttt{Kadath} syntax for differential equations. The explicit equations solved are not those given in Eqs.~\eqref{eq-phi}–\eqref{eq-W}, but the following compact system of elliptic PDEs,
\begin{subequations}\label{kadath_E}
\begin{eqnarray} 
\Delta_3 F_0  &= &4\pi A^2(\rho + S)
	+ \frac{B^2 r^2\sin^2 \theta}{2\alpha^2} \partial W\partial W  - \partial F_0 \partial(F_0 + F_2) \, ,\\
\tilde \Delta_3 (W r \sin\theta) &= &- 16\pi \frac{\alpha A^2}{B^2} \frac{P_\varphi}{r\sin\theta} + r\sin\theta  \, \partial W \partial(F_0 - 3 F_2) \, ,\\
\Delta_2 \left[ (\alpha B-1) r\sin\theta \right]
	&=& 8\pi \alpha A^2 B r\sin\theta (S^r_{\ \, r} + S^\theta_{\ \, \theta} ) \, ,\\
\Delta_2 (F_1 + F_0) &=& 8\pi A^2 S^\varphi_{\ \, \varphi} 
 + \frac{3 B^2 r^2\sin^2 \theta}{4 \alpha^2} \, \partial W\partial W - \partial F_0  \partial F_0  \, .
\end{eqnarray}
\end{subequations}
\begin{equation}\label{kadath_KG}
    \Delta_3  \phi - \frac{m^2}{r^2\sin^2\theta}\phi = A^2\left(\frac{dV}{d|\Phi|^2}-\frac{(\omega-mW)^2}{\alpha^2}\right)\phi - \partial\phi\partial(F_0+F_2) + \left(\frac{A^2}{B^2}-1\right)\frac{m^2}{r^2\sin^2\theta}\phi \, .
\end{equation}
Where we have introduced the notations  $\alpha = e^{F_0}$, $A = e^{F_1}$, and $B = e^{F_2}$ and the following differential operators:
\begin{subequations}\label{eq:operators}
\begin{eqnarray}
	&  & \Delta_2 := \partial_r^2 + \frac{1}{r}\partial_r
	+ \frac{1}{r^2}\partial_\theta^2 \, , \\
	& & \Delta_3 := \partial_r^2 + \frac{2}{r}\partial_r
	+ \frac{1}{r^2}\partial_\theta^2 + \frac{1}{r^2\tan\theta} \partial_\theta \, , \\
 	& & \tilde\Delta_3 := \Delta_3 - \frac{1}{r^2\sin^2\theta} \, , \\
    & & \partial f_1 \partial f_2 := \partial_r f_1 \partial_r f_2 + \frac{1}{r^2} \partial_\theta f_1 \partial_\theta f_2 \, .
\end{eqnarray}
\end{subequations}
The Klein–Gordon equation~\eqref{kadath_KG} is closely related to Eq.~\eqref{eq-phi}, differing only by a slight reformulation introduced to regularize the equation in the case of RBSs with $|m|=1$. More precisely, the term $m^2\phi/(r^2\sin^2\theta)$ is subtracted from both sides of the equation, a manipulation required to ensure regularity on the symmetry axis \cite{Grandclement:2014msa}.

The Einstein equations~\eqref{kadath_E} can be derived within the $3+1$ formulation of general relativity \cite{Bonazzola:1993zz}; a detailed derivation of the system can be found in \cite{Gourgoulhon:2010ju}. This formulation has proven well suited for high-precision spectral computations in the context of boson stars and related self-gravitating configurations, see e.g. \cite{Grandclement:2014msa,Garcia:2023ntf}. The source terms entering the Einstein equations correspond to the standard matter sources of the $3+1$ formalism and are obtained directly from the energy–momentum tensor in Eq.~\eqref{E-eq},
\begin{eqnarray}
    \rho + S &=& \frac{4}{\alpha^2}(\omega-mW)^2\phi^2 - 2\mu^2\phi^2 \, ,\\
    P_\varphi &=& 2\frac{(\omega-mW) m\phi^2}{\alpha} \, ,\\
    {S^r}_r + {S^\theta}_\theta &=& \frac{2}{\alpha^2}(\omega-mW)^2\,\phi^2 - 2\frac{m^2}{B^2r^2\sin^2\theta}\phi^2 - 2\mu^2\phi^2 \, ,\\
    {S^\varphi}_\varphi &=& \frac{(\omega-mW)^2\,\phi^2}{\alpha^2} + \frac{m^2}{B^2r^2\sin^2\theta}\,\phi^2 - \frac{\partial\phi\partial\phi}{A^2} - \mu^2\phi^2 ~~\, \, .
\end{eqnarray}
%
The nonlinear system is solved iteratively with \texttt{Kadath} constructing internally the Jacobian matrix associated with the spectral expansion coefficients of the fields. As noted in Sec.~\ref{spectral_Kadath}, boundary  
conditions are incorporated through the choice of the spectral basis functions\footnote{
In particular, the regularity conditions requiring vanishing $\theta$-derivatives of the metric functions at the axis, Eq.~\eqref{eq:derivative_pole_metric}, are satisfied identically by construction of the spectral basis and therefore do not need to be imposed separately.}, following the Galerkin spectral method. In the present implementation, all conditions described in Sec.~\ref{boundary} are imposed in this manner, with the sole exception of the no-conical singularity condition at the symmetry axis \eqref{eq:conic_sing}, $F_1=F_2$, which is not enforced, but verified \textit{a posteriori}, once a solution is obtained.

The computational domain is decomposed into several spherical spectral domains. Depending on the specific boson star configuration, different numbers of domains and different positions of the inter-domain boundaries are employed in order to improve accuracy. For each configuration, we specify explicitly in the code the number of domains, the radial position of the domain interfaces, and the number of spectral coefficients. In particular, the outermost domain employs\footnote{
The implementation of the different numerical radial coordinates is done internally; one does not need to rewrite the equations according to the specific coordinate transformation \cite{Grandclement:2009ju}.} a compactified radial coordinate extending to spatial infinity, whereas the inner domains use linear mappings of the radial coordinate $r$ that transform the interval between the domain boundaries, say, $r_a$ and $r_b$, onto the standard interval $[-1,1]$, where the Chebyshev polynomials are defined. The positions of $r_a$ and $r_b$ are chosen to adequately resolve regions with large scalar-field gradients. In all computations presented here, we choose the number of radial spectral coefficients to be equal to the number of angular spectral coefficients, $N_r = N_\theta \equiv N$, with the same value used across all domains, giving a total number of degrees of freedom 
\begin{equation}
    N_\text{DoF}=N^2\,n_\text{dom}\times N_K \, .
\end{equation}
The precise domain structure for each solution is summarized in Table~\ref{table_domains}. 
\begin{table}
    \centering
    \renewcommand{\arraystretch}{1.3}
    \begin{tabular}{ |c||c|c|l|  }
        \hline
        \textbf{Ref. solution} & $N$ & $n_{\rm dom}$ & location of boundaries \\
        \hline
        RBS1 & 21 & 8 & $2^n$ with $n\in[1,n_{\rm dom}-1]$\\
        RBS2 & 17 & 13& $2^n$ with $n\in[-4,n_{\rm dom}-6]$\\
        RBS3 & 21 & 10& $2^n$ with $n\in[-5,n_{\rm dom}-6]$\\
        DBS1 & 21 & 8 & $2^n$ with $n\in[1,n_{\rm dom}-1]$\\
        DBS2 & 17 & 9 & $2^n$ with $n\in[-1,n_{\rm dom}-3]$\\
        DBS3 & 33 & 9 & $2^n$ with $n\in[-1,n_{\rm dom}-3]$\\
        \hline
        \end{tabular}
    \caption{Number of spectral coefficients ($N$), number of spherical domains ($n_{\rm dom}$) and radial position of the boundaries for the reference solutions reported in Tables \ref{table_spin_newt}-\ref{table_dip_075}. Here, $N$ refers to $N_r$, the number of spectral coefficients per domain, which we choose to be equal to $N_\theta$, the number of angular coefficients. 
    }
    \label{table_domains}
\end{table}
%

Within each domain, the fields are expanded in spectral bases adapted to the geometry and parity properties of the solution. The number of spectral coefficients employed depends on the configuration considered and on the degree of localization of the scalar field. Typical resolutions explored in each case range from $N=9$ up to $N=25$ (for RBSs) or $N=33$ (for DBSs) coefficients per domain for the highest-accuracy solutions.
Given the exponential convergence property of spectral methods, smooth stationary BS configurations can be computed with high accuracy at comparatively moderate resolutions. 
This aspect of the code, with a relatively small number of degrees of freedom required for both moderate- and high-accuracy
solutions, was also discussed in Sec.~\ref{sec:runtime}.

Initial guesses are constructed from simple analytical profiles satisfying the correct asymptotic and regularity behavior. In practice, Gaussian-type ansätze for the scalar field, together with Minkowski values for the metric components, are sufficient to initialize the Newton iterations. For solution sequences, previously converged spectral solutions are used as initial data for nearby configurations. Specifically, the input parameter $F_0(r=0)$ is varied in steps corresponding to $\Delta\alpha(0)=\Delta e^{F_0(0)}=\pm0.01$, or smaller for highly relativistic boson stars and in the Newtonian regime. 
For the first configuration of a given family, the scalar-field trial functions used for the rotating and dipolar boson stars are taken from Refs.~\cite{Grandclement:2014msa,Jaramillo:2024cus}, which can be written as 
\begin{equation}
    \phi_{\rm RBS}=\phi_0(r\sin\theta)^m \,e^{-r^2\sin^2\theta/\sigma_x-r^2\cos^2\theta/\sigma_z} \, ,\qquad  \phi_{\rm DBS}=r\phi_0\cos\theta \,e^{-r^2\sin^2\theta/\sigma_x-r^2\cos^2\theta/\sigma_z},
    \label{eq:ig_expr}
\end{equation}
with $\phi_0$, $\sigma_x$ and $\sigma_z$ some constants that need to be tuned in order to converge toward an admissible solution. The remaining metric fields are initialized to their Minkowski-space values.  

Global quantities such as the ADM mass, angular momentum, Noether charge, and Komar integrals are computed directly within the \texttt{Kadath} scripts using the routines publicly available in the solver. In particular, asymptotic quantities are evaluated from the spectral representation of the fields at spatial infinity, while bulk quantities are obtained through multi-domain spectral integrations.
We have verified the exponential convergence of all relevant physical quantities when increasing the resolution. 
Explicit convergence tests are shown in Fig.~\ref{conv_kadath}, where the relative errors on global quantities and the virial identity exhibit the characteristic exponential decay with increasing spectral resolution before it saturates due to round-off errors. 
Unlike the convergence tests presented for \texttt{CADSOL} and \texttt{FreeFem}, here we plot the error estimators directly as functions of the spectral resolution $N$, rather than using the difference-based convergence estimator defined in Eq.~\eqref{eq:conv_estimator}. The reason is that, for the spectral basis employed by \texttt{Kadath}, the admissible numbers of radial spectral coefficients are restricted by the implementation \cite{Kadath} to values such as $N_r=9,11,13,17,21,25,33,\dots$. Consequently, the available resolutions are too sparse and non-uniform to construct a difference-based estimator, whereas they are sufficient to show the expected exponential convergence, following the same approach adopted in Refs.~\cite{Grandclement:2009ju,Grandclement:2014msa}.

As in the other approaches, extending the solver to fully 3D configurations within \texttt{Kadath} is relatively straightforward, since higher-dimensional spectral domains and tensorial operators are natively supported by the library. Indeed, related 3D elliptic problems, including numerical solutions of the Hamiltonian constraint involving BSs, have already been implemented within the \texttt{Kadath} framework in Ref.~\cite{Jaramillo:2024smx}. At the time of writing, the \texttt{Kadath} library\footnote{
Available from the repository \url{https://gitlab.obspm.fr/grandcle/Kadath.git}; documentation and examples can be found at \url{https://kadath.obspm.fr/}
} also includes, among its distributed examples, a full 3D numerical solution of the Kerr spacetime.


Finally, we note that several works on rotating spacetimes, particularly those employing \texttt{Kadath}, compute the ADM mass via a surface integral at spatial infinity involving the metric functions $F_1$ and $F_2$ (see Eq.~4.21 in \cite{Gourgoulhon:2010ju}), rather than $F_0$ as in Eq.~\eqref{eq:surf_adm}. We have verified that both definitions lead to the same conclusions regarding the relative difference with respect to the Komar volume integral. In this work, the surface integrals used to extract the ADM mass and angular momentum in \texttt{Kadath} are,
\begin{equation}
    M_\text{ADM}=\frac{1}{2}\lim_{r\to\infty}\int_0^{\pi}\partial_r \alpha \, r^2\sin\theta\,d\theta\, , \quad J_\text{ADM}=-\frac{1}{8}\lim_{r\to\infty}\int_0^\pi\partial_rW r^4 \sin^3\theta d\theta \, .
\end{equation}


\section{Future prospects: PINNs for rotating Q-balls}

\label{sec:future}


The recent explosion of machine learning and artificial intelligence has generated a great interest in the use of neural networks as a numerical method to solve differential equations \cite{Lee:1990, Meade:1994a, Meade:1994b, Yentis:1996, Lagaris:1997at, Lagaris:1997ap, Lagaris:2000, McFall:2009, Baymani:2010, Tarancon-Alvarez:2025wux}. We will focus here on a class of techniques usually known as physics-informed neural networks (PINNs) \cite{Raissi:2019, Nascimento:2020}. PINNs have been used in may areas of physics (see \cite{Desai:2021, Jin:2020, Mattheakis:2022, Bea:2024xgv, Jimenez:2026xhs, Ferrer:2024, Tseneklidou:2025stn} for a very incomplete list), including general relativity and the computation of quasinormal modes \cite{Ovgun:2019yor, Ncube:2021jfu, Cornell:2022enn, Luna:2022rql, Luna:2024spo, Patel:2024wzo, Cornell:2024azz, Ferrer-Sanchez:2026ubx}. The work in \cite{Liu:2026nve} is especially relevant for our case as it applies a semi-supervised approach to explore the families of spherically symmetric, non-rotating BSs. They first generate BS families with a finite-element Newton solver, then train using pointwise supervision from those solutions together with the EKG equation residuals and global constraints on the 
quantities $M$ and $Q$. We present here a complemetary approach 
which is completely unsupervised, i.e., it does not rely on previously generated data and instead finds the solution using only the field equations.

The basic idea behind PINNs is to use a simple fully connected neural network as a universal approximant \cite{Hornik:1989} for a mathematical function that must be fitted in order to satisfy a set of differential equations. In this sense, the neural network plays the role of the interpolants that are used in more standard methods, i.e., Taylor series in finite differences, or a basis of functions for spectral or finite-element methods. Finite-difference methods use Taylor's theorem to relate the function derivatives to the function values in neighboring points. Spectral-like methods use some orthogonality relation in the basis to compute the coefficients that interpolate the solution. 

However, in the case of PINNs we do not have a closed-form mathematical recipe to compute the free coefficients of the approximant. Instead, we have to optimize them iteratively using a given quantitative criterion that measures how far the network is from the desired behavior. This is encoded in a loss function, which has to be designed in such a way that it decreases as we approach the solution. Then, we typically use an optimization algorithm to minimize the loss. This is achieved by some variant of stochastic gradient descent (SGD). In this paper we will use Adam for the first physics-informed iterations and L-BFGS for the final iterations, as we will explain later. We have implemented the whole pipeline in the standard machine learning framework \texttt{PyTorch}.

\begin{figure}[ht]
\begin{center}
    \includegraphics[width=0.99\linewidth]{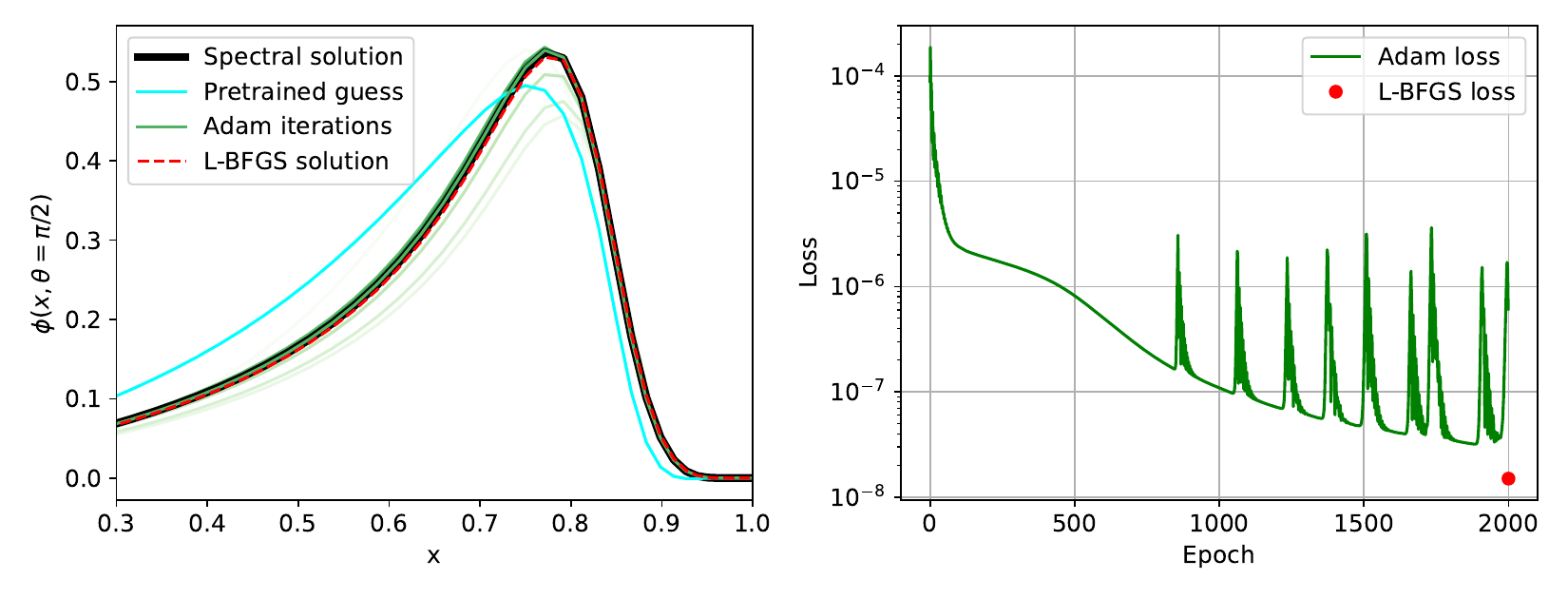}
    \caption{Left: Scalar field profile on the equatorial line $\phi(x, \theta = \pi / 2)$ during the three phases of the PINN training: pretrained guess, Adam iterations and L-BFGS final result. Different shades of green correspond to different epochs, with light green corresponding to early stages and dark green to late stages. We plot the spectral solution as a reference, showing that it is very close to the PINN result. Right: Evolution of the physics-informed loss during the Adam training epochs, and after L-BFGS polishing.}
    \label{fig:pinn_iterations}
\end{center}
\end{figure}

\begin{figure}[h!]
\begin{center}
    \includegraphics[width=0.99\linewidth]{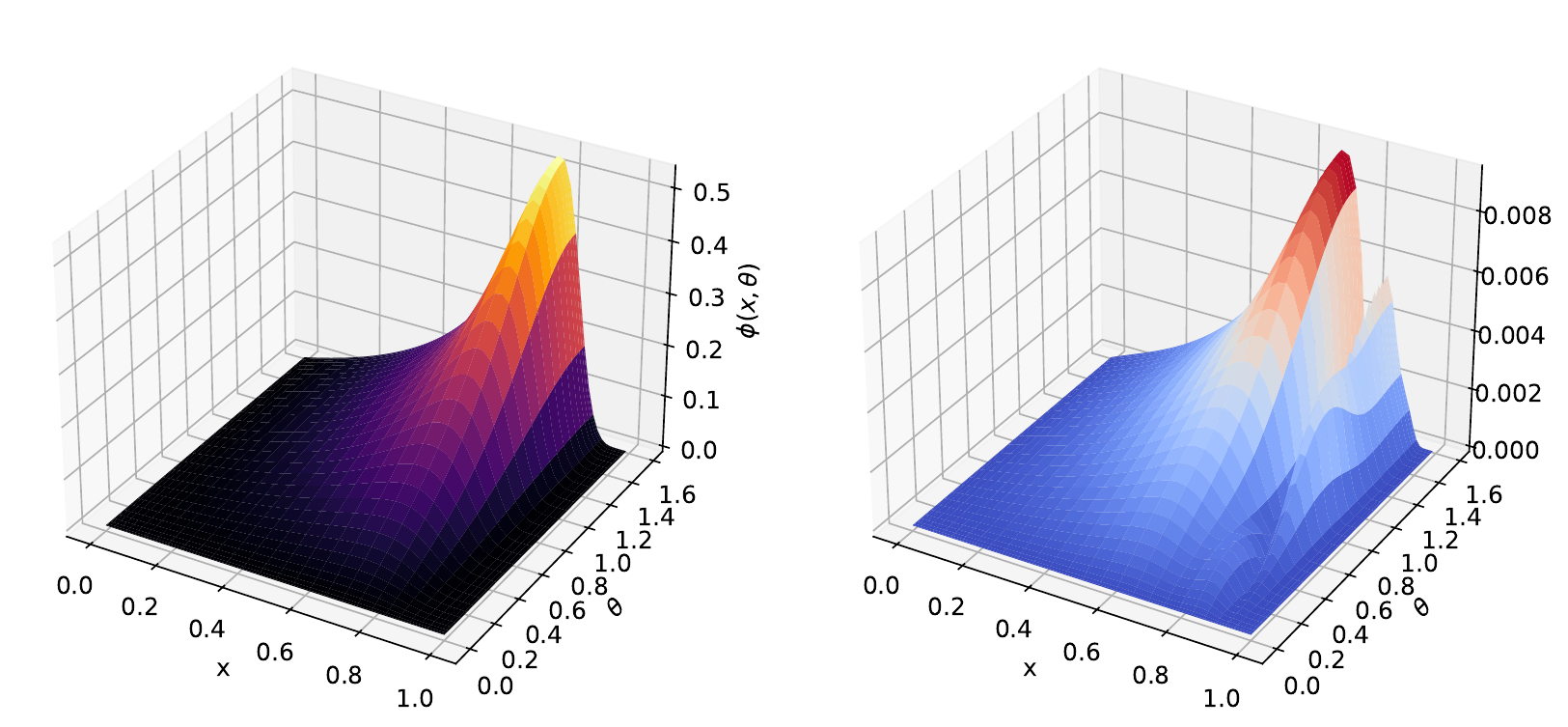}
    \caption{Left: Scalar field profile in compactified coordinates $(x, \theta)$ as obtained from the PINN training process. Right: Absolute value of the difference between the PINN result and the reference spectral solution, with a maximum difference of 0.0093 on the $64 \times 62$ spectral grid nodes.}
    \label{fig:pinn_solution}
\end{center}
\end{figure}

When considering 
the full coupled EKG system to look for BS solutions, minimizing the PINN loss has a strong tendency to converge towards the trivial flat-space, zero-scalar solution. Avoiding this effect requires additional constraints and training strategies which would make the present exercise considerably more involved. Therefore, as a proof of concept in the context of this paper, we provide here a simple example for a Q-ball solution \cite{Volkov:2002aj, Kleihaus:2005me}. A Q-ball is a non-topological solitonic configuration of a self-interacting scalar field. It can be viewed as the flat space analogue of a BS, with scalar self-interactions providing the binding mechanism instead of gravity. In this case, the absence of a dynamical spacetime simplifies the problem significantly, which makes it ideal as a toy model. 
We therefore consider a variant of the action (\ref{action}) in Minkowski spacetime, with a potential term $U(\left|\Psi\right|^2 )$,
\begin{equation}
    S=\int d^4 x \left[ - \frac12 g^{\mu\nu} \left( \Psi_{, \, \mu}^* \Psi_{, \, \nu} + \Psi_{, \, \nu}^* \Psi _{, \, \mu} \right) - U(\left|\Psi\right|^2 ) \right].
\label{eqn:qball_action}
\end{equation}
We will focus on a potential of the form \cite{Volkov:2002aj, Kleihaus:2005me},
\begin{equation}
    U( \left|\Psi\right|^2 ) = \lambda |\Psi|^2 \left(|\Psi|^4 - a |\Psi|^2 + b \right) ,
\label{eqn:qball_potential}
\end{equation}
and look for axisymmetric solutions, using the scalar ansatz (\ref{scalar_ansatz}), which results in a simplified form of (\ref{eq-phi}), now with a potential term,
\begin{equation}
    \left( \frac{\partial^2} {\partial r^2} + \frac{2}{r} \frac{\partial}{\partial r} + \frac{1} {r^2} \frac{\partial^2} {\partial \theta^2} + \frac{\cos\theta} {r^2 \sin\theta} \frac{\partial} {\partial \theta} - \frac{m^2} {r^2 \sin^2\theta} + \omega^2 \right) \phi = U'(\phi)\, , 
\label{eqn:qball_PDE}
\end{equation}
where we take $U(\phi) = \lambda \phi^2 \left(\phi^4 - a \phi^2 + b \right)$. For the example in this paper, we will use the physical parameters $\lambda = 0.5, \, a = 2, \, b = 1.1, \, m = 1, \, \omega = 0.99$. We will also use the radial compactification from (\ref{comp}) with $c = 1$, from which the differential equation can be cast in the form,
\begin{equation}
    \Delta_Q \phi = S(\phi)\, ,
    \label{eqn:qball_eom_simple}
\end{equation}
with the differential operator $\Delta_Q$ and source $S(\phi)$ term defined as,
\begin{equation}
    \Delta_Q\phi = x(1-x)^4\sin^2\theta\,(x\,\partial_x^2\phi+2\,\partial_x\phi) + (1-x)^2\sin^2\theta\,\partial_\theta^2\phi + (1-x)^2\sin\theta\cos\theta\,\partial_\theta\phi\, ,
\end{equation}
\begin{equation}
    S(\phi) = (1-x)^2 m^2\phi - x^2\sin^2\theta \left(\omega^2\phi - U'(\phi)\right)\, .
\end{equation}
The boundary conditions for $\phi$ are the same as for rotating BSs, see Sec.~\ref{boundary}. Since automatic differentiation is natively implemented in \texttt{PyTorch}, we can use the ``hard enforcement'' of such boundary conditions without any extra coding effort. In contrast to ``weak enforcement'', where boundary conditions are imposed by adding extra terms to the loss function, ``hard enforcement'' satisfies the boundary conditions exactly by designing the neural network interpolant appropriately. In our case, we take the approximated field to be,
\begin{equation}
    \phi_\text{model}(x,\theta) = x e^{- \kappa r} \sin \theta \, \mathcal{N} (x, \cos^2\theta)\,, 
\end{equation}
where $\mathcal{N} (x, \cos^2\theta)$ is a neural network with two input neurons and one output neuron. The network has 4 hidden layers of 64 neurons (12737 total trainable parameters), with hyperbolic tangent activation functions. Notice that the prefactor of $ x e^{- \kappa r} \sin \theta$ automatically imposes the Dirichlet boundary conditions $\phi = 0$ whenever $x = 0, 1$ or $\theta = 0$. Additionally, feeding the neural network the pair $(x, \cos^2 \theta)$ instead of just $(x, \theta)$ is a straightforward way to impose Neumann boundary conditions at 
$\theta = \pi / 2$. The particular choice of $e^{- \kappa r}$, with 
\begin{eqnarray}
    \kappa = \sqrt{U''(0) - \omega^2} = \sqrt{2 \lambda b - \omega^2} \, ,
\end{eqnarray}
ensures that the field will decay at the correct rate as $r \to \infty$. This is especially important when computing the total energy $E$ and charge $Q$, 
as they strongly depend on the asymptotic behavior of the solution \cite{Volkov:2002aj}. The training process is divided in three major phases:
\begin{itemize}
    \item {\bf Phase 1: Pretraining of an ansatz:} As any iterative method, PINNs often require an initial guess before we start converging to a solution. This guess can be a previous solution with slightly different physical parameters, or an analytic ansatz. In our case, we use the double Gaussian profile, 
    \begin{equation}
        \phi_\text{guess}(x, \theta) = \frac12 e^{- \tfrac18 \left[(z_c)^2+(x_c-3)^2 \right]} - \frac12 e^{- \tfrac18 \left[(z_c)^2+(x_c+3)^2 \right]} \, ,
    \end{equation}
    where the Cartesian coordinates $x_c$ and $z_c$ are computed as usual,
    \begin{equation}
        z_c = \frac{x}{1-x} \cos \theta, \qquad x_c = \frac{x}{1-x} \sin \theta\, .
    \end{equation}
    The fit is performed using 2000 epochs of an Adam optimizer with a learning rate of $10^{-3}$. The loss function is the mean squared error (MSE) over a 50 $\times$ 50 grid $(x_i, \theta_j)$, linearly spaced in both directions, as
    \begin{equation}
        \mathcal{L}_1 = \frac{1}{N_x N_\theta} \sum_{i, j} \left[ \phi_\text{model}(x_i, \theta_j) - \phi_\text{guess}(x_i, \theta_j) \right]^2\, .
    \end{equation}
    After the pretraining, we obtain a loss of $\left. \mathcal{L}_1 \right|_\text{final} = 8.423 \times 10^{-7}$. If we assume that our ansatz is not too far from the the solution we want, this is already a test of the capability of the neural network to fit its general features. Notice that the differential equations have not been used yet.
    \item {\bf Phase 2: Physics-informed training:} At this point we introduce the differential equation (\ref{eqn:qball_eom_simple}) into the scheme, 
    to improve the initial guess of Phase 1 towards a solution. Here we still use 2000 epochs of an Adam optimizer with a learning rate of $10^{-3}$, but in this phase the loss is physics-informed. 
    We use as a loss function the mean squared residuals from (\ref{eqn:qball_eom_simple}), evaluated on the grid points, 
    \begin{equation}
    \label{eq:L2}
        \mathcal{L}_2 = \frac{1}{N_x N_\theta} \sum_{i, j} \left[\Delta_Q \phi - S(\phi) \right]^2_{x_i, \theta_j}\, .
    \end{equation}
    Here we use a 32 $\times$ 32 tensor product Chebyshev-Lobatto grid, which should help prevent overfitting. However, the method can technically be used with an arbitrary set of points with no modification. Figure \ref{fig:pinn_iterations} shows the loss $\mathcal{L}_2$ progression as a function of the training epoch, as well as a representation of the field profile as it approaches the optimal solution. The final loss value after this phase is $\left. \mathcal{L}_2 \right|_\text{final} = 6.102 \times 10^{-7}$.

    \item {\bf Phase 3: Polishing by L-BFGS:} Even though the loss from phase 2 is very small, the solution is not necessarily very accurate. For this reason, once we are in the right basin of attraction, keeping the same loss function~\eqref{eq:L2}, we apply a final single epoch of a Limited-memory Broyden–Fletcher–Goldfarb–Shanno (L-BFGS) optimizer. Notice that the \texttt{PyTorch} implemantation of L-BFGS has internal iterations for each optimization epoch, in this case we set a maximum of 800. This is a more sophisticated optimizer that uses an estimate of the Hessian matrix to take into account the curvature of the loss before updating the neural network's parameters. We use the \texttt{PyTorch} implementation of L-BFGS with the following parameters: \texttt{lr=1.0}, \texttt{max\_iter=800}, \texttt{history\_size=100}, \texttt{tolerance\_grad=1e-15}, \texttt{tolerance\_change=1e-17}, \texttt{line\_search\_fn=``strong\_wolfe''}. We use a denser 56 $\times$ 56 tensor product Chebyshev-Lobatto grid. This final polishing step is also shown in Figure \ref{fig:pinn_iterations}. The final loss has dropped to $\left. \mathcal{L}_3 \right|_\text{final} = 1.499 \times 10^{-8}$, and the solution profile has become closer to the solution.
\end{itemize}
In order to assess the accuracy of the PINN, we compare it with a reference solution that has been obtained using collocation spectral methods on a Chebyshev grid, via Newton-Raphson iterations. Figure \ref{fig:pinn_iterations} shows the scalar field profile on the equatorial line $\phi(x, \theta = \pi / 2)$ along the process, starting from the pretrained guess, then the Adam iterations and finally the L-BFGS polished solution. The spectral solution is also plotted as a reference. Figure \ref{fig:pinn_solution} shows the final result from the PINN, and its difference from the reference solution. We also compare two integrated global quantities, namely, the energy $E$ and charge $Q$, defined in Ref.~\cite{Volkov:2002aj} as,
\begin{equation}
    E = 2\pi\int_0^\infty dr\, r^2 \int_0^\pi d\theta\,\sin\theta\left(\omega^2\phi^2+(\partial_r\phi)^2+\frac{1}{r^2}(\partial_\theta\phi)^2+\frac{m^2\phi^2}{r^2\sin^2\theta}+U(\phi)\right)\, ,
    \label{eqn:qball_energy}
\end{equation}
\begin{equation}
    Q = 4\pi\omega\int_0^\infty dr\,r^2\int_0^\pi d\theta\,\sin\theta\,\phi^2\, .
    \label{eqn:qball_charge}
\end{equation}
The derivatives in Eq.~(\ref{eqn:qball_energy}) are obtained by automatic differentiation on the trained PINN, and by Chebyshev differentiation matrices on the spectral solution. Then the integration on the compactified domain is performed with high precision using Clenshaw-Curtis weights. Table \ref{tab:EQ_pinn_comparison} shows the values obtained on the PINN solution, compared to the reference. The relative error typically falls in the 1-2\% level.
\begin{table}[ht]
\centering
\begin{tabular}{|l|c|c|}
\hline
 & $E$ & $Q$ \\ \hline
Spectral & 136.06 & 157.09 \\ \hline
PINN & 137.83 & 159.64 \\ \hline
\% error & 1.30 & 1.63 \\ \hline
\end{tabular}
\caption{Comparison of energy $E$ and charge $Q$ from the PINN and the reference spectral solution.}
\label{tab:EQ_pinn_comparison}
\end{table}

\section{Final Remarks}

\label{sec:conclusion}

  
The main purpose of this work was to present a comparative analysis of different numerical methods for constructing exotic compact objects in relativistic gravity, taking BSs as a well-established testbed. We focused on the simplest \textit{non-spherical} configurations: rotating BSs and static dipolar BSs. The solutions were obtained using three independent codes based respectively on finite differences (\texttt{FIDISOL/CADSOL}), finite elements (\texttt{FreeFem}), and spectral methods (\texttt{Kadath}). Overall, excellent agreement was found across the six references configurations, with the values of mass, angular momentum (or Noether charge), and lapse at the origin agreeing to 8 significant digits in most cases, and up to 9 in some. The most challenging case is RBS3, which lies on the third branch of the mass-frequency spiral and corresponds to a strongly localized scalar field configuration; here the agreement decreases to 6 digits. These results represent the highest precision ever reported for physical quantities of BSs, with the typical errors in the existing literature being of order $10^{-4}$ (see \textit{e.g.} Ref.~\cite{Cunha:2022tvk} on DBS), whereas the values reported here have errors indicators between $10^{-9}$ and $10^{-15}$. 

\medskip

The numbers provided in this work may therefore serve as reliable benchmark data for future numerical studies of BSs and for the development of new numerical frameworks. This will be particularly valuable in the era of artificial intelligence-based methods such as PINNs, which are attracting growing interest for the numerical resolution of nonlinear PDEs such as those arising in strong-gravity problems. In particular, supervised learning methods require a large catalog of solutions to be available during the training phase of the model. An excellent example of this is Ref.~\cite{Liu:2026nve}, which shows clearly that high-precision conventional solvers remain necessary both for generating the training/reference configurations and for certifying individual solutions. Standard PINNs can be unsupervised, but hybrid physics-informed/surrogate approaches to BSs may rely directly on catalogs generated by conventional numerical codes.

\medskip

Each code has its own strengths and limitations. 
The main advantage of \texttt{FIDISOL/CADSOL} consists in its flexibility. It can be installed on any platform and has a simple, intuitive structure that makes it easy to become familiar with.
Moreover, interfacing with symbolic computation packages such as \textit{Mathematica} is straightforward, enabling a coherent interactive “notebook” workflow.
A specific feature of \texttt{FIDISOL/CADSOL} is that, in addition to the unknown functions, their first and second derivatives are also directly computed inside the solver.
Also, it does not require a highly accurate initial guess for the solutions, and has therefore been successfully used to solve a wide range of field theory problems.
At the same time, this code is no longer supported, and neither the software nor most of the documentation is publicly available.
Also, for problems with a larger number of functions and configurations with steep gradients, it has proven  difficult to obtain high accuracy results, mesh adaptation or manual mesh design being necessary.
Moreover, as seen in Table \ref{table_runtime},  when considering the runtime to obtain a given solution,
\texttt{FIDISOL/CADSOL} is  less efficient, as compared to other two solvers in this work.

The \texttt{FreeFem}-based code proves particularly effective for configurations that are numerically demanding in terms of spatial resolution, such as the strongly localized cases RBS3 and DBS3 and the very dilute cases RBS1 and DBS1. In all these cases, the high flexibility of finite elements in terms of mesh adaptivity allows for targeted refinement near the origin or the asymptotic boundary -- where the field profiles exhibit steep gradients -- with non-uniform meshes straightforwardly implemented. The main practical limitation of the code is that it requires the \texttt{FreeFem} library and all its dependencies to be installed; comprehensive documentation, including installation instructions, are  however available online. The code developed for this work is also publicly available on GitHub~\cite{Gervalle_Spinning_mini-Boson_Stars_2025} and, while designed specifically for the BSs model~\eqref{action}, its structure is sufficiently modular to be adapted to other field theory models with relative ease. We also note that the finite element implementation developed here is based exclusively on first-order derivatives, which is standard in the FEM approach. Computing second derivatives is technically possible but not recommended, as their evaluation with this solver tends to introduce significant numerical noise.

The \texttt{Kadath}-based code benefits from the exponential convergence typical of the spectral approach -- as opposed to the power-law convergence of finite difference and finite element methods -- allowing excellent accuracies 
to be reached with very moderate resolutions. A further advantage is that both the spectral infrastructure and the BS solver are included in the public release of the library. Moreover, the symbolic syntax used to specify the field equations is intuitive, making it straightforward to modify the existing solver or implement new stationary configurations if similar boundary and regularity conditions are required.
For the solutions considered in this work, the main limitation is the saturation of the spectral convergence coming from round-off errors. In practice, this limits the attainable accuracy to approximately $10^{-10}-10^{-9}$; achieving higher accuracies such as $10^{-13}$ 
does not appear to be hindered by insufficient spectral resolution or computational resources, but rather by aspects of the numerical implementation itself, possibly including the evaluation of global quantities. As with \texttt{FreeFem}, another practical limitation is the set of external dependencies required by the library, which may make its installation on computing clusters technical. 

Beyond the comparison of different numerical codes and approaches, this work has also made progress on an open question regarding DBS. While the parameter space of RBS 
 is better understood -- with (presumably) infinitely many branches
of solutions spiraling toward a limiting configuration at the center of the mass-frequency diagram -- the corresponding picture for DBS was unclear. Indeed, while a minimal frequency associated with the first back-bending of the $(\omega,M)$ curve is observed, the study of Ref.~\cite{Cunha:2022tvk} reported only part of the second branch of solutions, which ended around ${\omega/\mu \simeq 0.79}$  due to the increasing numerical errors.
Using the \texttt{FreeFem}-based code, we have been able to extend this second branch and reach the second back-bending at ${\omega/\mu \simeq 0.84284}$ (see left panel of Fig.~\ref{fig_DBS}). No sign of pathological behavior is observed along this branch; however, the numerics become increasingly challenging, with error estimates reaching $\sim10^{-4}$ for the last computed solution and the required frequency steps $\Delta\omega$ needed to maintain convergence becoming smaller and smaller, ultimately preventing further extension of the sequence. We argue that these difficulties arise as the scalar field distribution becomes extremely localized at a finite distance along the symmetry axis, which is insufficiently resolved in the current implementation. While a complete understanding of the limiting behavior for DBSs remains outstanding, our results are consistent with the usual spiral picture. Its full resolution would require a more involved mesh adaptivity strategy, which we leave for future work.

\section*{Acknowledgements}

R.G. would like to thank Julien Garaud for useful discussions throughout the project and for sharing his expertise in \texttt{FreeFem}, in particular regarding the parallelization strategy. E.S.C.F. would like to thank Angelo Guimarães for helpful discussions on elliptic partial differential equations. This work is supported  by the  Center for Research and Development in Mathematics and Applications (CIDMA) through the Portuguese Foundation for Science and Technology (FCT -- Fundac\~ao para a Ci\^encia e a Tecnologia), references  UIDB/04106/2020 and UIDP/04106/2020.  
The authors acknowledge support  from the projects CERN/FIS-PAR/0027/2019, PTDC/FIS-AST/3041/2020,  CERN/FIS-PAR/0024/2021 and 2022.04560.PTDC.  
This work has further been supported by  the  European  Union's  Horizon  2020  research  and  innovation  (RISE) programme H2020-MSCA-RISE-2017 Grant No.~FunFiCO-777740 and by the European Horizon Europe staff exchange (SE) programme HORIZON-MSCA-2021-SE-01 Grant No.~NewFunFiCO-101086251. Computations have been performed at the Argus and Blafis cluster at the U. Aveiro. This work was supported by FCT I.P. under Project 2025.09655.CPCA.A2 através da FCCN at Deucalion supercomputer, jointly funded by EuroHPC JU and Portugal. E.S.C.F. is partially supported by the FCT grant PRT/BD/153349/2021 (\url{https://doi.org/10.54499/PRT/BD/153349/2021}) under
the IDPASC Doctoral Program and by CNPq/PDJ 153723/2025-4.
V.J. acknowledges support from Secretaría de Ciencia, Humanidades, Tecnología e Innovación (SECIHTI, México) through a postdoctoral fellowship.
%
We acknowledge financial support from the ``Center of Excellence Maria de Maeztu 2025–2029'' award to the ICCUB, grant CEX2024-001451-M, funded by MICIU/AEI/10.13039/501100011033 and  from Grant No. PID2022-136224NB-C22 from the Spanish Ministry of Science, Innovation and Universities, and from Grant No. 2021-SGR-872 funded by the Catalan Government. This research is also funded by the European Union (ERC, HoloGW, Grant Agreement No. 101141909). Views and opinions expressed are, however, those of the authors only and do not necessarily reflect those of the European Union or the European Research Council. Neither the European Union nor the granting authority can be held responsible for them.

\section*{AI and LLMs usage disclaimer}
Standard AI programming tools were used during the development of the code that yielded the results shown in this paper. These were used to increase efficiency in tasks such as debugging and modifying existing code. Large language models were used to assist in editing and polishing the manuscript, but they did not contribute to the scientific content or results.

	\bibliographystyle{hhieeetr}
	\bibliography{biblio}
	
\newpage

\end{document}